\documentclass[twocolumn,floats,preprintnumbers,superscriptaddress,a4,nofootinbib,longbibliography,aps]{revtex4-2}

\usepackage{kotex}
\usepackage{graphicx}
\usepackage{bm}
\usepackage{color}
\usepackage{hyperref}
\usepackage[normalem]{ulem}
\usepackage{amsmath}
\usepackage{amssymb}
\usepackage{mathtools}
\usepackage{here}
\usepackage[dvipsnames]{xcolor}

\usepackage{multirow}

\newcommand{\figidx}[1]{(#1)}

\newcommand{\ie}{{{i.e.}}}
\newcommand{\eg}{{{e.g.}}}

\newcommand{\kT}{k_{B} T}
\newcommand{\enat}{\text{e}}
\newcommand{\deri}[2]{\frac{{d}{#1}}{{d}{#2}}}

\def\Equation{Equation}

\def\Eq{Eq.}
\def\Eqs{Eqs.}

\def\Fig{Fig.}
\def\Figs{Figs.}
\def\Tab{Table}

\extrafloats{100}

\begin{document}

\title{Transport in polymer membranes beyond linear response:\\ Controlling permselectivity by the driving force}

\author{Won Kyu Kim}
\email{wonkyukim@kias.re.kr}
\affiliation{Korea Institute for Advanced Study, Seoul 02455, Republic of Korea}

\author{Sebastian Milster}
\affiliation{Research Group for Simulations of Energy Materials, Helmholtz-Zentrum Berlin f\"ur Materialien und Energie, D-14109 Berlin, Germany}
\affiliation{Applied Theoretical Physics-Computational Physics, Physikalisches Institut, Albert-Ludwigs-Universit\"at Freiburg, D-79104 Freiburg, Germany}

\author{Rafael Roa}
\affiliation{Departamento de F\'{i}sica Aplicada I, Facultad de Ciencias, Universidad de M\'{a}laga, E-29071 M\'{a}laga, Spain}

\author{Matej Kandu\v{c}}
\affiliation{Jo\v{z}ef Stefan Institute, SI-1000 Ljubljana, Slovenia}

\author{Joachim Dzubiella}
\email{joachim.dzubiella@physik.uni-freiburg.de}
\affiliation{Research Group for Simulations of Energy Materials, Helmholtz-Zentrum Berlin f\"ur Materialien und Energie, D-14109 Berlin, Germany}
\affiliation{Applied Theoretical Physics-Computational Physics, Physikalisches Institut, Albert-Ludwigs-Universit\"at Freiburg, D-79104 Freiburg, Germany}
\affiliation{Cluster of Excellence livMatS @ FIT - Freiburg Center for Interactive Materials and Bioinspired Technologies, Albert-Ludwigs-Universit\"at Freiburg, D-79110 Freiburg, Germany}

\date{\today}

\begin{abstract}
In the popular solution-diffusion picture, the membrane permeability is defined as the product of the partition ratio and the diffusivity of penetrating solutes inside the membrane in the linear response regime, i.e., in equilibrium.  However, of practical importance is the penetrants' flux across the membrane driven by external forces.   Here, we study nonequilibrium membrane permeation orchestrated by a uniform external driving field using molecular computer simulations and continuum (Smoluchowski) theory in the stationary state. In the simulations, we explicitly resolve the penetrants' transport across a finite monomer-resolved polymer network, addressing one-component penetrant systems and mixtures. We introduce and discuss possible definitions of nonequilibrium, force-dependent permeability, representing `system' and `membrane' permeability. In particular, we present for the first time a definition of the {\it differential permeability} response to the force. We demonstrate that the latter turns out to be significantly nonlinear for low-permeable systems, leading to a high amount of selectiveness in permeability, called `permselectivity', and is tunable by the driving force.  Our continuum-level analytical solutions exhibit remarkable qualitative agreement with the penetrant- and polymer-resolved simulations, thereby allowing us to characterize the underlying mechanism of permeabilities and steady-state transport beyond the linear response level.  
\end{abstract}


\maketitle

\section{introduction}

Solvated polymer membranes are in general made from cross-linked polymer networks (e.g., hydrogels). 
They are omnipresent in different fields such as soft matter, bio- and materials sciences. 
Living systems are replete with polymer networks, such as cytoskeletons, mucus, and extracellular matrices. They often function as selective walls (barriers) that control transport of various biomolecules~\cite{shasby1982role,wingender1999bacterial,hay2013cell,witten2017particle,goodrich2018enhanced,fuhrmann2020diffusion,joo2020anomalous}. 
From an engineering perspective as well, synthetic polymer networks are very important: They are indispensable building blocks, serving as core components in dyalisis, nanofiltration and desalination~\cite{reverse}, or in emerging `smart' materials, such as drug delivery systems~\cite{peppas1999, stamatialis, moncho2020scaling} and stimuli-responsive  nanoreactors~\cite{CarregalRomero:2010gp,stuart2010emerging,Lu:2011bi,Vriezema:2005fx,Renggli:2011if,Tanner:2011jf,Gaitzsch:2015kr,Guan:2011eva,Herves:2012fp,Wu:2012bx,Prieto:2016bj,Petrosko:2016he,Jia:2016cy,quesada2012computer} that comprise catalysts with sophisticated functionality inside a spherical hydrogel shell. These nanoreactors bear the potential for diverse applications, starting from well-controlled chemical reactions ~\cite{CarregalRomero:2010gp,stuart2010emerging,Lu:2011bi,Vriezema:2005fx,Renggli:2011if,Tanner:2011jf,Gaitzsch:2015kr,Guan:2011eva,Herves:2012fp,Wu:2012bx,Prieto:2016bj,Petrosko:2016he,Jia:2016cy} to biomedical diagnoses~\cite{Vriezema:2005fx,Guan:2011eva,Renggli:2011if,Tanner:2011jf,Gaitzsch:2015kr}, where various membrane materials, ranging from artificial polyelectrolytes to biomacromolecules, are utilized~\cite{Liu:2016gf,Montolio:2016jy,Zinchenko:2016jy,Erbas2015}. 

In particular, the transport in liquid solutions involves molecular solutes, embracing nano-scale atoms to sub-micron-scale macromolecules, such as ions, ligands, proteins, and reactants, which need to penetrate dense and solvated membranes. An important quantity to characterize such a solute transport through the membrane is the {\it permeability}. The latter has developed over the last decades to become the central material property describing the nature of transport phenomena in the natural sciences and engineering~\cite{graham1866lv,missner2009110,al1999one, finkelstein1987water,al1999one,venable2019molecular,WOLDEKIDAN2021463}.
Particularly important for applications is to utilize solute selectivity in the permeability (`permselectivity') of polymeric membranes of different morphologies, as, \eg, prominently found in air filtration or gas separation~\cite{robeson, chauhan, MOF, Lyd, Obliger, freeman1999basis,park2017,bilchak2020tuning}, and water purification~\cite{shannon:nature,geise2010water,geise2011,Menne2014,tansel,tan2018polyamide,hyk2018water,lu2019tuning}.

\begin{figure}[t!]
\centering
\includegraphics[width = 0.38\textwidth]{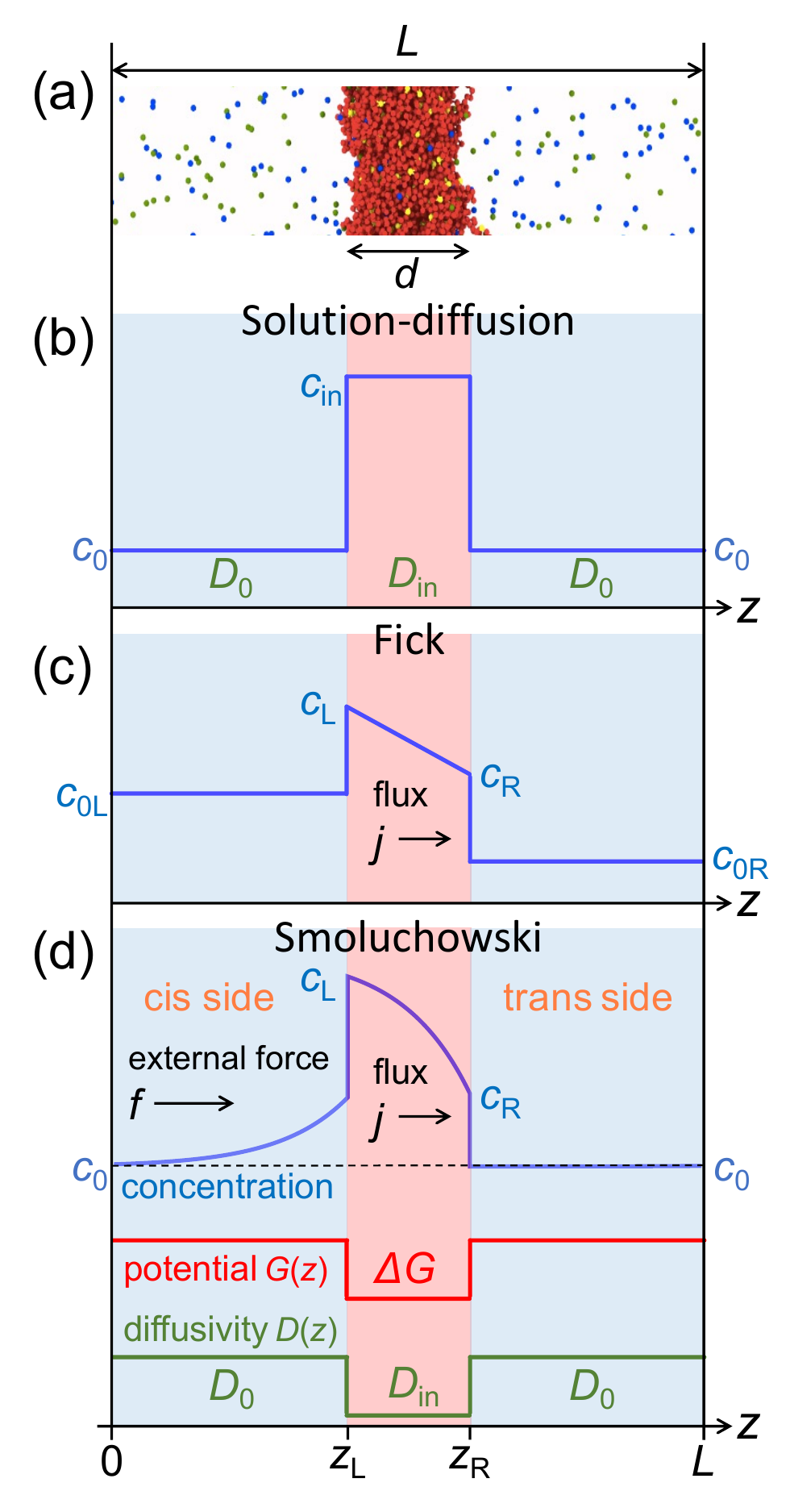}
\caption{Various scenarios of membrane permeation. \figidx{a} A polymer network membrane of thickness $d$ (red), located at the center of a simulation box of length $L$, in equilibrium with the penetrants (small blue and green spheres).   
\figidx{b} A continuum version of system (a): solution-diffusion model with equilibrium penetrant concentrations $c_0$ and $c_{\rm in}$ in bulk and inside the membrane, respectively, and corresponding diffusion coefficients $D_0$ and $D_{\rm in}$.
\figidx{c}~The Fick's type permeation: The flux $j$ is generated by different bulk reservoir concentrations of penetrant solutes $c_\text{0L}$ and $c_\text{0R}$.
\figidx{d}~The Smoluchowski type permeation in nonequilibrium: The flux $j$ is generated by an external force $f$ acting on penetrants, which flows from the cis side to the trans side. $G(z)$ and $D(z)$ are the position-dependent potential and diffusivity, respectively [\eg,~see \Eqs~\eqref{eq:Gz} and \eqref{eq:Dz}].
}\label{fig:models}
\end{figure}

Early in history~\cite{meyer1899eigenschaft,overton1901studien,missner2009110}, it was argued  that the membrane permeability ($\mathcal{P}$) is proportional to partitioning (also referred to as the partition coefficient or the partition ratio) $\mathcal{K} \equiv c_\text{in}/c_{0}$, that is, the ratio of penetrant concentrations inside the membrane ($c_\text{in}$) and a bulk reservoir ($c_{0}$).  
The permeability relation has been further studied and extended for dense membranes, and the solution-diffusion model was established, leading to~\cite{yasuda1969permeability3,diamond1974interpretation,paul1976solution,robeson,Baker, finkelstein1987water,gehrke, mason1990statistical,palasis1992permeability,chauhan,al1999one,Thomas2001transport,Ulbricht2006,missner2009110,Baker2014,venable2019molecular,WOLDEKIDAN2021463} 
\begin{equation}\label{eq:P}
\mathcal{P} \equiv  D_\text{in} \mathcal{K},
\end{equation}
which is the product of the two key quantities, $\mathcal{K}$ and $D_\text{in}$, where the latter is the diffusion coefficient of the penetrants inside the membrane.\footnote{Some works use a different definition $\mathcal{P} \equiv  D_\text{in} \mathcal{K}/d$ for the permeability. Here we consider $\mathcal{P} \equiv  D_\text{in} \mathcal{K}$, thereby independent of the membrane thickness $d$.} The solution-diffusion model is based on the linear response assumption, \ie, it describes the permeability in terms of equilibrium quantities, as illustrated in \Figs~\ref{fig:models}\figidx{a} and \figidx{b}. The permeability can be interpreted as the penetrant diffusivity weighted by the partitioning.  Therefore, the permeability of the bulk reservoir is simply the penetrant's free bulk diffusivity, $\mathcal{P}_0=D_0$. 

In fact, the permeability defined in \Eq~\eqref{eq:P} is the proportionality constant of the flux $j$,
\begin{equation}\label{eq:j_ficks}
j \equiv -\mathcal{P} \frac{\Delta c_0}{d},
\end{equation}
driven by the penetrant concentration difference $\Delta c_0 = c_\text{0R}-c_\text{0L}$ between both sides of a membrane of thickness $d$ (the Fick's type law)~\cite{gehrke,nelson2004biological,venable2019molecular}. 
We illustrate this situation of membrane permeation in \Fig~\ref{fig:models}\figidx{c}.
The relation, \Eq~\eqref{eq:j_ficks}, may be useful to measure $\mathcal{P}$ in experiments where penetrant bulk concentrations on both sides are kept constant but are different under the complete stirring condition [$c_\text{0L}$ and $c_\text{0R}$ in \Fig~\ref{fig:models}\figidx{c}].
Combining Fick's law with the partitioning $\mathcal{K}$, one finds $j=-D_\text{in}(c_\text{R}-c_\text{L})/d=-D_\text{in}\mathcal{K}(c_\text{0R}-c_\text{0L})/d$, which derives the solution-diffusion definition for $\mathcal{P}$ shown in \Eq~\eqref{eq:P}. Note that the flux is always proportional to the concentration gradient (i.e., we are still in the linear response regime). 

A more general approach to derive the permeability is utilizing the steady-state Smoluchowski equation (or the Fokker-Planck equation for overdamped dynamics~\cite{risken1996fokker,doi1988theory}) of the penetrant particles in the presence of an external driving force $f$. In this framework, the flux is generally described as
\begin{equation}\label{eq:Pj}
j = - D(z) \left[ \deri{c(z)}{z} + c(z) \beta \left( \deri{G(z)}{z} - f\right) \right],
\end{equation}
including contributions of the potential $G(z)$ and diffusivity $D(z)$ landscapes (due to the membrane), and where $\kT=1/\beta$ is the thermal energy. 
The proposed approach generalizes the inhomogeneous solution-diffusion model (\ie, with position-dependent potential and diffusion coefficient)~\cite{diamond1974interpretation} in the presence of a uniform, external force. 
The corresponding situation is relevant to measurements of permeability utilizing an external driving force beyond the linear response level.
In biophysics, such an `active' transport~\cite{albers1967biochemical,baranowski1991non} of solutes (not to be confused with the self-propulsive `active particles'~\cite{ramaswamy2010mechanics,romanczuk2012active,PhysRevLett.80.5044,PhysRevLett.99.048102}) stems from externally applied forces (\eg, by carrier proteins in cell membranes~\cite{alberts2002carrier,bonifacino2003coat}) to generate a flux of solutes even against concentration gradients.  Other examples of such force-driven transport are experiments that generate a flux by solvent flow as in reverse osmosis or driven by an electric field as in electrodialysis~\cite{reverse}. 
 
In fact, the stationary Smoluchowski picture based on \Eq~\eqref{eq:Pj} has been assessed as a fundamental theoretical framework used in a multitude of physical problems to ascertain the underlying mechanism of stochastic barrier-crossing processes under a uniform, time-invariant external field. Some pioneering examples include diffusion in tilted periodic potentials~\cite{stratonovich1965oscillator,costantini1999threshold,reimann2001giant,lindner2001optimal}, polymer translocation~\cite{sung1996polymer}, Josephson tunneling junctions~\cite{ambegaokar1969voltage,anchenko1969josephson}, and molecule separations using electrophoresis~\cite{ajdari1991free}, to name a few~\cite{risken1996fokker,reimann2002brownian,burada2009diffusion,hanggi2009artificial}. However, a discussion on membrane permeability in this framework is considerably rarely addressed. Moreover, a comprehensive study is not even fully conducted, particularly for nonequilibrium permeability under a driving force.

\begin{figure}[t]
\centering
\includegraphics[width = 0.47\textwidth]{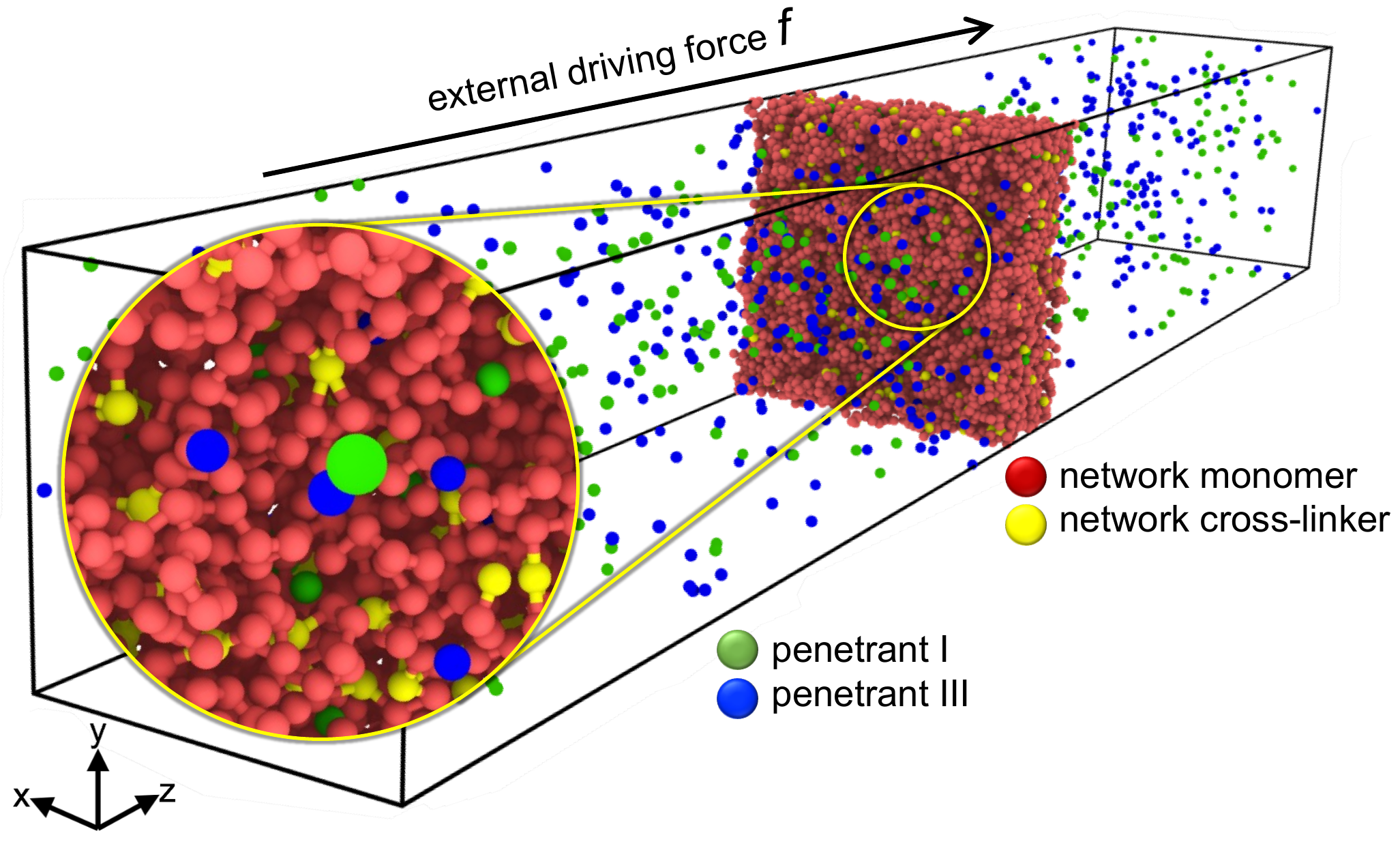}
\caption{Simulation snapshot of a polymer network membrane in the presence of two types of diffusive penetrants (type I in green, type III in blue; type II not included in this simulation) driven by an external force $f$ in the $z$-direction (arrow). The polymers (red) are polydisperse in length and connected by tetra-functional cross-linkers (yellow). The magnification circle shows the membrane structure in more detail.
}\label{fig:snapshot}
\end{figure}

In this work, we aim at connecting and contrasting the `classical' equilibrium solution-diffusion relation with `driven' transport in the framework of the foregoing Smoluchowski equation.
The theoretical prediction is supplemented by coarse-grained (CG) molecular simulations of a (monomer-resolved) polymer network membrane. The goal also extends our previous theoretical work for equilibrium permeability~\cite{Rafa2017,kim2019prl,kim2020tuning} to nonequilibrium scenarios. We seek reasonable definitions of force-dependent permeabilities in the system out of equilibrium, in which physical quantities in consideration reduce to equilibrium ones in the force-free limit. In particular, we investigate to what extent the driving force can selectively control the the permeability of one penetrant species to the other, when the system comprises penetrants of different kinds (cosolutes).
Separating or filtrating a specific type of molecules in multi-component mixtures is an indispensable process. 
In this context, selectivity is an essential quantity to determine a measure of relative permeation across a membrane.
To this end, we carry out CG simulations of one- and two-component-penetrant transport through a membrane of polymer networks in the presence of an external driving force. The system consists of a randomly formed polymer network membrane~\cite{kim2020tuning} in contact with a bulk reservoir~\cite{kim2017cosolute,kim2019prl}, as illustrated in \Figs~\ref{fig:models}\figidx{a} and \ref{fig:snapshot}. The solute penetrants thus diffuse in both bulk regions under the driving force. We demonstrate that our theoretical predictions correspond with the simulation results for this nonequilibrium model membrane transport. Importantly, we discuss how permselectivity can be tuned by the driving force which provides experimentalist with another means to control solute transport.   

\section{Simulation model and Theory}

\subsection{Simulation model}

We perform Langevin dynamics simulations of cross-linked, semi-flexible polymer networks in the presence of diffusive solutes (penetrants) driven by an external  force, as illustrated in \Fig~\ref{fig:snapshot}.
The polymers form a random tetra-functional network membrane in which the cross-linker fraction is around 5\%~\cite{kim2020tuning}.
The membrane is tethered at the center of the simulation box with harmonic constraints over the entire simulation time. The details of the membrane morphology, simulation setup, protocols, and CG force field parameters can be found in Methods as well as in our previous work on equilibrium permeability~\cite{kim2020tuning}.

Briefly, for nonbonded interactions we consider Lennard-Jones (LJ) potentials $U_\text{LJ}^{ij}$ for $i, j =$ n or p, where n denotes the network particles (chain monomers and cross-linkers), and p stands for the penetrants. 
For the interactions between the penetrants we use the LJ strength $\beta\epsilon_\text{pp}=0.1$~\cite{kim2017cosolute,kim2019prl,kim2020tuning}, at which the short-range and steep repulsion of the LJ potential dictates the interaction (steric exclusion limit). The intra-network interaction $\epsilon_\text{nn}$ is interpreted as a measure of solvent quality~\cite{Heyda2013,kim2017cosolute,kim2019prl,kim2020tuning} (thus controls the network volume fraction $\phi_\text{n}$).
We use $\beta\epsilon_\text{nn}=0.5$ in all our simulations, which
yields $\phi_\text{n} \approx 0.2$ in equilibrium ($f=0$). Another core parameter that strongly tunes the permeability is the network-penetrant interaction $\beta\epsilon_\text{np}$~\cite{kim2020tuning}.
In this work, we consider three different penetrant types (termed $\text{p}_\text{I}$, $\text{p}_\text{II}$, and $\text{p}_\text{III}$) for which we use $\beta\epsilon_{\text{n}\text{p}_\text{I}}=0.1$, $\beta\epsilon_{\text{n}\text{p}_\text{II}}=0.6$, and $\beta\epsilon_{\text{n}\text{p}_\text{III}}=1.2$, as shown in \Tab~\ref{tab:para}.
For single-component penetrant systems, we consider each penetrant type per system.
For the mixture of two-component penetrants  (see \Fig~\ref{fig:snapshot}), we consider two different parameter sets to probe the selective behavior of the membrane, namely `Mixture 1' of $\{\text{p}_\text{I}, \text{p}_\text{III}\}$ and `Mixture 2' of $\{\text{p}_\text{II}, \text{p}_\text{III}\}$.
The unit length $\sigma$ is used in the model, which is the particle diameter identical for all particles.
We summarize all parameters used in this study in \Tab~\ref{tab:para}.

\begin{table}
\caption{\label{tab:para} Parameter values for the network-penetrant interaction strength $\beta \epsilon_\text{np}$, equilibrium membrane permeability $\mathcal{P}^\text{eq}$, penetrant inner-membrane diffusivity $D_\text{in}$, partitioning $\mathcal{K}$ and the equilibrium membrane thickness $d/\sigma$, for three different penetrants ($\text{p}_\text{I}$, $\text{p}_\text{II}$, and $\text{p}_\text{III}$), where $D_0$ is the penetrant free diffusivity in the bulk reservoir. The system's longitudinal length $L=305\sigma$ with the unit length $\sigma$ (monomer diameter), the network-network interaction parameter $\beta\epsilon_\text{nn}=0.5$, and the penetrant-penetrant interaction parameter $\beta\epsilon_\text{pp}=0.1$ are fixed for all systems in consideration. We consider three different single-component penetrant systems with each penetrant type. We also consider two sets of two-component penetrant mixtures, namely `Mixture 1' of $\{\text{p}_\text{I},\text{p}_\text{III}\}$ and `Mixture 2' of $\{\text{p}_\text{II},\text{p}_\text{III}\}$.}
\begin{ruledtabular}
\begin{tabular}{llllll}
Penetrant type & $\beta \epsilon_\text{np}$ & $\mathcal{P}^\text{eq}/D_0$ & $D_\text{in}/D_0$ & $\mathcal{K}$ & $d/\sigma$ \\ \hline         
$\text{p}_\text{I}$ & 0.1 & 0.13 & 0.36 & 0.36 & 14.2\\
$\text{p}_\text{II}$ & 0.6 & 0.40 & 0.26 & 1.54 & 14.6\\
$\text{p}_\text{III}$ & 1.2 & 3.30 & 0.13 & 25.4 & 22.4\\
\end{tabular}
\end{ruledtabular}
\end{table}

We apply a constant (in both space and time) force $f$ to all penetrants in the $z$-direction (see the arrow in \Fig~\ref{fig:snapshot}) for a sufficiently long time, ensuring a steady state where the penetrants' flux $j(z,f)$ is independent of $z$, \ie, constant [see the Supporting Information (SI)].
For varying values of $f$, we analyze time-averaged penetrant concentration profiles $c(z,f)$, velocity profiles $v_z(z,f)$ in the $z$-direction, and thus the mean flux $j(f)=\langle c(z,f)v_{z}(z,f)\rangle$ as main quantities to describe the force-dependent permeability and selectivity.
Here $\langle X \rangle = \int dz X /L$ denotes the additional average over the longitudinal space, where $L$ is the system length in the $z$-direction.

\subsection{Theory}

\subsubsection{One-dimensional steady-state solutions of the Smoluchowski equation}

To connect to continuum transport theory, we assume that the simulation model can be projected into one dimension (1D) along the $z$-direction as illustrated in \Fig~\ref{fig:models}, and that penetrants behave as an ideal gas in a 1D energy landscape. 
This is owing to the symmetry in the lateral $xy$-direction, and vanishing penetrant-penetrant interactions except the excluded volume in the simulations.
Hence, we consider ideal pointlike penetrants in a total potential $U(z,f)=G(z) - f z$ in 1D with an external constant force $f$ and a position-dependent diffusivity $D(z)$. 
We recall the steady-state Smoluchowski equation introduced in \Eq~\eqref{eq:Pj} from which 
the steady-state flux $j$ with boundary conditions $c(z_1)=c_1$ and $c(z_2)=c_2$ yields the constant flux~\cite{stratonovich1965oscillator,risken1996fokker},
\begin{equation}
\label{eq:flux}
j = \frac{{c_1 \enat^{\beta U(z_1,f)} - c_2 \enat^{ \beta U(z_2,f)}}}{I(z_1,z_2,f)},
\end{equation}
with 
\begin{equation}
\label{eq:Iz}
I(z_1,z_2,f) \equiv \int_{z_1}^{z_2} dy \frac{\enat^{\beta U(y,f)}}{D(y)}.
\end{equation}
The general solution for the position- and force-dependent penetrant concentration leads to,
\begin{eqnarray}
\label{eq:sol2}
c(z,f) &=& \enat^{-\beta U(z,f)}\left[ c_1 \enat^{\beta U(z_1,f)} - \right. \nonumber \\
&&\left.{\left({c_1 \enat^{\beta U(z_1,f)} - c_2 \enat^{ \beta U(z_2,f)}}\right)}\frac{I(z_1,z,f)}{I(z_1,z_2,f)} \right].
\end{eqnarray}

\subsubsection{Flux, Fick's law, and linear response}

The flux obtained in \Eq~\eqref{eq:flux} is a nonlinear function of $f$. Here, we briefly discuss the nature of $j$ and its connection to Fick's law, the solution-diffusion linear response, and permeability.
To this end, consider a membrane-only system with a constant potential ($\Delta G$) and constant diffusivity ($D_\text{in}$) in the range of $z_\text{L} \leq z \leq z_\text{R}=z_\text{L}+d$ (\ie, a membrane of thickness $d$). This yields the total potential $U(z,f)=\Delta G-fz$, and one finds from \Eq~\eqref{eq:flux} the exact expression
 \begin{eqnarray}\label{eq:j_example}
j = D_\text{in} {\beta f} \frac{c(z_\text{L}) - c(z_\text{R}) \enat^{- \beta f d}}{1-\enat^{-\beta f d}},
\end{eqnarray}
where the inner-membrane boundary concentrations of penetrants, $c(z_\text{L})$ and $c(z_\text{R})$, are arbitrary. Expressing in terms of a Taylor series for $f$, it reads
\begin{eqnarray}\label{eq:j_example2}
j &=& - \frac{D_\text{in} \Delta c}{d}+\frac{1}{2} D_\text{in} \left \{ c(z_\text{L})+c(z_\text{R}) \right\} \beta f \nonumber\\
&&-\frac{d}{12} D_\text{in} \Delta c (\beta f)^2+\Delta c~\mathcal{O}\left(f^4\right),
\end{eqnarray}
with the concentration difference denoted by $\Delta c  = c(z_\text{R})-c(z_\text{L})$.
The first term (zeroth order of $f$) reveals Fick's law that defines the solution-diffusion equilibrium membrane permeability [see \Eq~\eqref{eq:j_ficks}] with $f=0$, where the flux is driven by $\Delta c$ in nonequilibrium. The second term indicates the linear relation of $j$ and $f$, which is proportional to the arithmetic mean of the boundary concentrations, $[c(z_\text{L})+c(z_\text{R})]/2$. 
We learn from the above expansion that the flux is generated by two sources, $\Delta c$ and $f$.

Note that for vanishing $\Delta c$, the second term in \Eq~\eqref{eq:j_example2} is the only nonvanishing term.
This implies that when the penetrant's inner-membrane boundary concentrations are $c(z_\text{L})=c(z_\text{R}) \equiv c_\text{in}$ in equilibrium, the nonlinear flux, $j$ in \Eq~\eqref{eq:j_example}, explicitly reduces to the linear function
\begin{eqnarray}\label{eq:j_example4}
j &=& D_\text{in} c_\text{in} \beta f,
\end{eqnarray}
meaning that the flux inside the membrane is described by the equilibrium quantities $D_\text{in}$ and $c_\text{in}$ as a linear response to $f$.

{\subsubsection{Definition of a nonequilibrium system permeability}

By using the equilibrium membrane permeability defined in \Eq~\eqref{eq:P}, for which from now on we denominate $\mathcal{P}^\text{eq} = D_\text{in}c_\text{in}/c_0$ with superscript `eq', one finds from \Eq~\eqref{eq:j_example4}
\begin{eqnarray}\label{eq:j_def}
j = \mathcal{P}^\text{eq} c_0 \beta f.
\end{eqnarray}
This connects $j$ with the penetrant concentration in a reference reservoir ($c_0$) via $\mathcal{P}^\text{eq}$ (or partitioning therein). 
\Equation~\eqref{eq:j_def}, which we explicitly derive here, provides an important $j$-$f$ relation that defines $\mathcal{P}^\text{eq}$ as the proportionality factor on the linear response level, whereas the solution-diffusion model defines $\mathcal{P}^\text{eq}$ in $j$-$\Delta c$ relation.

Motivated by the above linear response derived from the membrane-only system, we extend the notion to a nonequilibrium situation beyond the linear response.
In this case the system consists of a membrane immersed in the reference bulk reservoir of solutes [see \Figs~\ref{fig:models}\figidx{d} and \ref{fig:snapshot}].
A steady-state flux, $j(f)$, measured in this system is constant throughout space and time, regardless of being measured in the membrane or in the bulk.  
We thus bring forward the system's total permeability, $\mathcal{P}_\text{sys}$, defined in a general fashion as proportionality factor of the system's observable $j(f)$ via
\begin{equation}\label{eq:j_Psys}
j(f) \equiv \mathcal{P}_\text{sys}(f) c_0 \beta f.
\end{equation}

This inspires us to define a force-dependent {\it chordal or mean} system permeability,  
\begin{eqnarray}\label{eq:P_j}
\mathcal{P}_\text{sys}(f) \equiv \frac{j(f)}{c_0 \beta f},
\end{eqnarray}
which shares the same philosophy with the force-dependent mobility $\mu(f) = v(f)/f$ defined as the apparent prefactor of the velocity $v$ with respect to the force~\cite{risken1996fokker,dagdug2012force}.
In addition, we consider another quantity for permeation, defining the {\it incremental or differential} ~system permeability via
\begin{eqnarray}\label{eq:increP}
\mathcal{P}^\Delta_\text{sys}(f) \equiv \frac{1}{\beta c_0} \frac{dj(f)}{df},
\end{eqnarray}
which is analogous to the definition of differential conductivity in charge transport in other systems, such as semiconductors or lattice gases~\cite{esaki1970superlattice,lei1991theory,labouvie2015negative}. It describes the change of the flux with respect to isothermal incremental changes of the driving force.
Both quantities, $\mathcal{P}_\text{sys}$ and $\mathcal{P}^\Delta_\text{sys}$, coincide in the force-free limit ($f \rightarrow 0$), \ie, they are independent of $f$ (linear response).  However, they can be force-dependent and different in higher force regimes in general. 

In this work, we differentiate $\mathcal{P}_\text{sys}(f)$ and $\mathcal{P}^\Delta_\text{sys}(f)$ by terming the former simply `system permeability' and the latter explicitly `differential system permeability'. 
In addition, we denominate $\mathcal{P}^\text{eq}$, $\mathcal{P}(f)$, and $\mathcal{P}^\Delta(f)$ for `equilibrium membrane permeability', `membrane permeability', and `differential membrane permeability', respectively. 
Note that these definitions define a total system permeability, not the individual membrane permeability in contact with a solute reservoir (as in equilibrium), because they are based on the global flux which is determined not only by membrane properties but also system size and boundary conditions.  We comment on and discuss a possible definition of a nonequilibrium membrane permeability [$\mathcal{P}(f)$ and $\mathcal{P}^\Delta(f)$] later in section III.C.

\begin{figure*}[t]
\centering
\includegraphics[width = 0.93\linewidth]{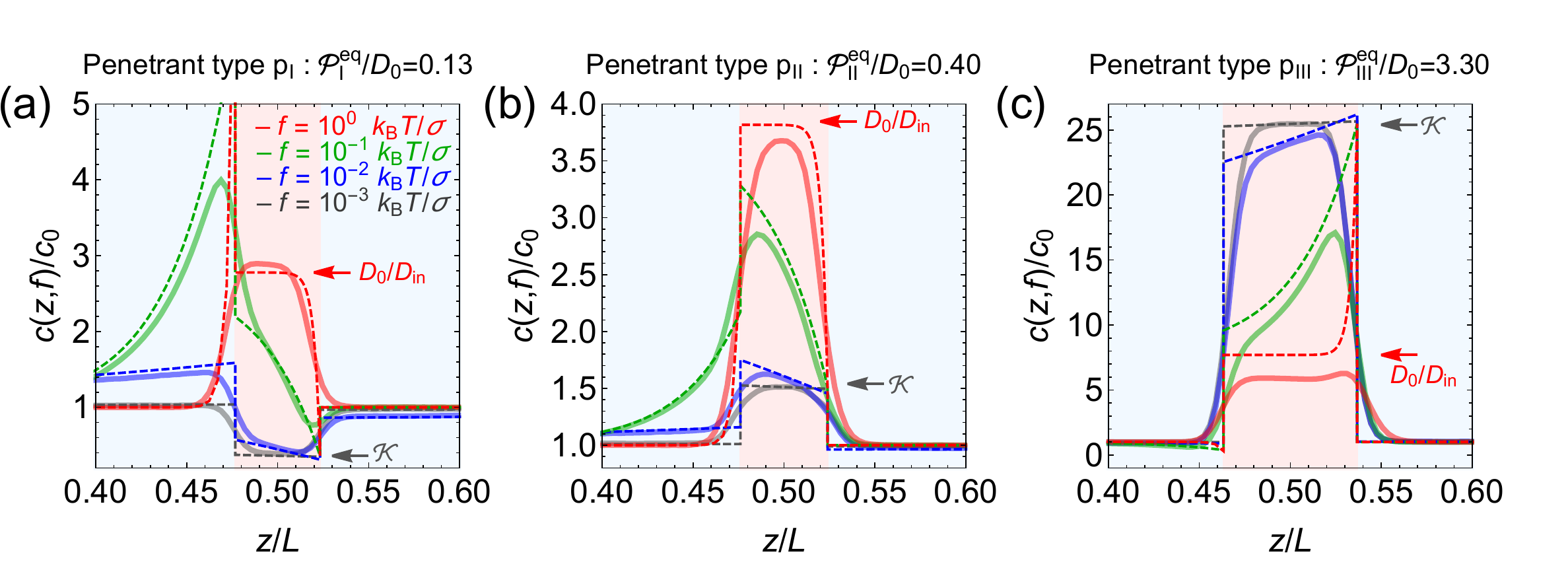}
\caption{
Normalized penetrant concentration profiles $c(z,f)/c_0$ in single-component penetrant systems for different forces. The red shaded area depicts the membrane region with corresponding equilibrium membrane thickness $d$. The time-averaged steady-state profiles from the simulation results (solid lines) and the theoretical prediction (dashed lines) from the steady-state Smoluchowski solution [\Eq~\eqref{eq:sol_full}] are compared for different parameters \figidx{a}~$\beta\epsilon_{\text{np}_\text{I}} = 0.1$ and $\mathcal{P}_\text{I}^\text{eq}/D_0=0.13$, \figidx{b}~$\beta\epsilon_{\text{np}_\text{II}} = 0.6$ and $\mathcal{P}_\text{II}^\text{eq}/D_0=0.4$, and \figidx{c}~$\beta\epsilon_{\text{np}_\text{III}} = 1.2$ and $\mathcal{P}_\text{III}^\text{eq}/D_0=3.3$. The arrows depict the limiting values for the penetrant's inner-membrane concentration, $c_\text{in}/c_0=\mathcal{K}$~for $f\rightarrow 0$ [gray, see \Eq~\eqref{eq:cf0}], and $c_\text{in}(L/2)/c_0=D_0/D_\text{in}$~for $f\rightarrow \infty$ [red, see \Eq~\eqref{eq:cfinf}]. The used parameter values are summarized in \Tab~\ref{tab:para}.}
\label{fig:3}
\end{figure*}

\section{Results and discussion}

\subsection{Single-component penetrant system}

Equipped with the general steady-state solutions and definitions, let us now consider a special system of a membrane with a bulk reservoir of penetrants [see \Fig~\ref{fig:models}\figidx{d}], for which we assume a piecewise constant form for both the potential and the diffusivity, according to 
\begin{equation}\label{eq:Gz}
G(z)=
\begin{cases}
\Delta G &z_\text{L} \leq z \leq z_\text{R}, \\
0 &\text{elsewhere},
\end{cases}
\end{equation}
and
\begin{equation}\label{eq:Dz}
D(z)=
\begin{cases}
D_\text{in} &z_\text{L} \leq z \leq z_\text{R}, \\
D_0 &\text{elsewhere},
\end{cases}
\end{equation}
with $z_\text{L}=L/2-d/2$ and $z_\text{R}=L/2+d/2$.
The penetrants are of one kind (single-component), driven by an external force $f$, and have the same bulk concentrations at both system boundaries $c(0)=c(L)=c_0$. 
The total potential energy then is $U(z,f)=G(z)-fz$, thus $U(0,f)=0$ and $U(L,f)=-fL$.
The solution $c(z)$ conforms to the imposed boundary condition for $G(z)$ and $D(z)$, which is periodic in the $z$-direction with the period $L$~\cite{risken1996fokker}. This enables the comparison with the simulation results in which the periodic boundary conditions are used.

\subsubsection{Penetrant concentration profiles}

It is straightforward to obtain the penetrant concentration profile from the general solution \Eq~\eqref{eq:sol2}, reading 
\begin{equation}
\label{eq:sol_full}
\frac{c(z,f)}{c_0}= \left[1 -  \left( 1- \enat^{-\beta f L}\right)\frac{I(0,z,f)}{I(0,L,f)} \right] \enat^{-\beta U(z,f)}.
\end{equation}
The full closed-form expression of \Eq~\eqref{eq:sol_full} is presented in the Appendix, where the penetrant concentrations are denoted by the following subscripts depending on the location,  
\begin{equation}
c(z,f)=
\begin{cases}
c_\text{0L}(z,f) &0 \leq z < z_\text{L}, \\
c_\text{in}(z,f) &z_\text{L} \leq z \leq z_\text{R}, \\
c_\text{0R}(z,f) &z_\text{R} < z \leq L. \nonumber
\end{cases}
\end{equation}

In \Fig~\ref{fig:3}, we compare $c(z,f)/c_0$ from the simulation results (solid lines) with the theoretical prediction [\Eq~\eqref{eq:sol_full}, dashed lines] for the single-component-penetrant-membrane systems. The standard error of the time-averaged concentration profiles obtained from the simulations is smaller than the line thickness.
We visualize the membrane region by depicting the red shaded areas with corresponding equilibrium membrane thickness $d$ (see \Tab~\ref{tab:para} and Methods). 
We notice at first glance that the particle-based simulation results are in remarkable agreement with the continuum-level theory based on the piecewise potential model, overall for different forces and membrane network-penetrant interaction parameters $\epsilon_\text{np}$.

For the repulsive network-penetrant interactions with the penetrant type I ($\beta\epsilon_{\text{np}_\text{I}} = 0.1$) yielding the smallest equilibrium membrane permeability ($\mathcal{P}_\text{I}^\text{eq}/D_0=0.13$) in the chosen parameter set, the change of $c(z,f)$ upon increasing $f$ is significant, as shown in \Fig~\ref{fig:3}\figidx{a}. Since the interaction is repulsive, the global magnitude of the penetrant's inner-membrane concentration [$c_\text{in}(z,f)$] is smaller than in the reservoir bulk regions [$c_\text{0L}(z,f)$ and $c_\text{0R}(z,f)$], thus $\mathcal{K}<1$, for small forces up to $\beta f=0.01/\sigma$ close to equilibrium. As the force further increases, a dramatic inversion of $c(z,f)$ occurs, in which the penetrants infiltrate into the membrane much more than outside, eventually resulting in the increase of the membrane permeability (will be discussed in a subsequent section). In particular, under the intermediate force, $\beta f=0.1/\sigma$, a giant penetrant accumulation around the left (entry or cis~\cite{sung1996polymer}) side of the membrane-reservoir interface is revealed, which even exceeds overall the inner concentration $c_\text{in}(z,f)$, as depicted by the green line in \Fig~\ref{fig:3}\figidx{a}. This induces a rapidly decaying concentration profile $c_\text{in}(z,f)$ that becomes extremely inhomogeneous in this intermediate force regime, which decays even below the bulk concentration at the right (exit or trans) side of the membrane.

For the slightly attractive network-penetrant interactions with the penetrant type II ($\beta\epsilon_{\text{np}_\text{II}}  = 0.6$) yielding $\mathcal{P}_\text{II}^\text{eq}/D_0=0.4$ but still less than unity [see \Fig~\ref{fig:3}\figidx{b}], the penetrant's interfacial accumulation around the cis side is observed again under the intermediate driving force (the green line). However, there is no dramatic inversion of $c_\text{in}(z,f)$ with respect to the bulk concentrations as $f$ varies. This tendency is related to the partitioning, that is, $\mathcal{K}>1$ in this case. The penetrant's inner-membrane concentration tends to increase as $f$ increases. Similarly to the results in \Fig~\ref{fig:3}\figidx{a}, the inner concentrations $c_\text{in}(z,f)$ decay with a negative gradient, $d c_\text{in}(z,f)/dz<0$. 

The above observed cis-side interfacial accumulation implies a penetrants' congestion due to the intrinsically low permeability of the membrane, \ie, low $D_\text{in}$ and low $\mathcal{K}$.
This jamming at the entry is released by exceeding driving forces that compensate for the penalty from the low permeability, and eventually becomes negligible under higher forces. 
For example, in the case of \Fig~\ref{fig:3}\figidx{a}, the maximal cis-side accumulation occurs at $\beta f=0.1/ \sigma$. The potential barrier height in this case is $\beta\Delta G = -\ln \mathcal{K} \approx 1$.
This is comparable to the energy needed to move a penetrant over a distance of the membrane thickness ($d/\sigma\approx14$) with $\beta f=0.1/ \sigma$, which amounts to $\beta f d \approx 1.4$.
Thus, there occurs a large interfacial accumulation. 
The interfacial accumulation is released with the larger force $\beta f=1/ \sigma$ that estimates the corresponding membrane-crossing energy $\beta f d \approx 14$, which is over tenfold larger than $\beta\Delta G$.

For the more attractive network-penetrant interactions with the penetrant type III ($\beta\epsilon_{\text{np}_\text{III}}  = 1.2$) yielding $\mathcal{P}_\text{III}^\text{eq}/D_0=3.3$ and now larger than unity [see \Fig~\ref{fig:3}\figidx{c}], we find a significantly different behavior of $c(z,f)$. Firstly, owing to the negative energy barrier arising from the membrane, there is no salient interfacial accumulation of the penetrants. Secondly, the magnitude of $c_\text{in}(z,f)$ decreases with $f$, which reduces partitioning but the flux increases. Lastly, now  $c_\text{in}(z,f)$ increases with $z$, thereby having a positive gradient $d c_\text{in}(z,f)/dz>0$.  
This positive gradient signifies a flux to the $z$-direction against the concentration gradient, which is often found in biophysics, featuring active transport~\cite{albers1967biochemical,baranowski1991non}.
The sign change of the concentration gradient depends on $\mathcal{P}^\text{eq}$. We find the threshold of this sign change at $\mathcal{P}^\text{eq}=D_0$, which is evident from $d c_\text{in}(z,f)/dz \propto (\mathcal{P}^\text{eq}-D_0)$.

\begin{figure*}[t]
\centering
\includegraphics[width = 0.75\linewidth]{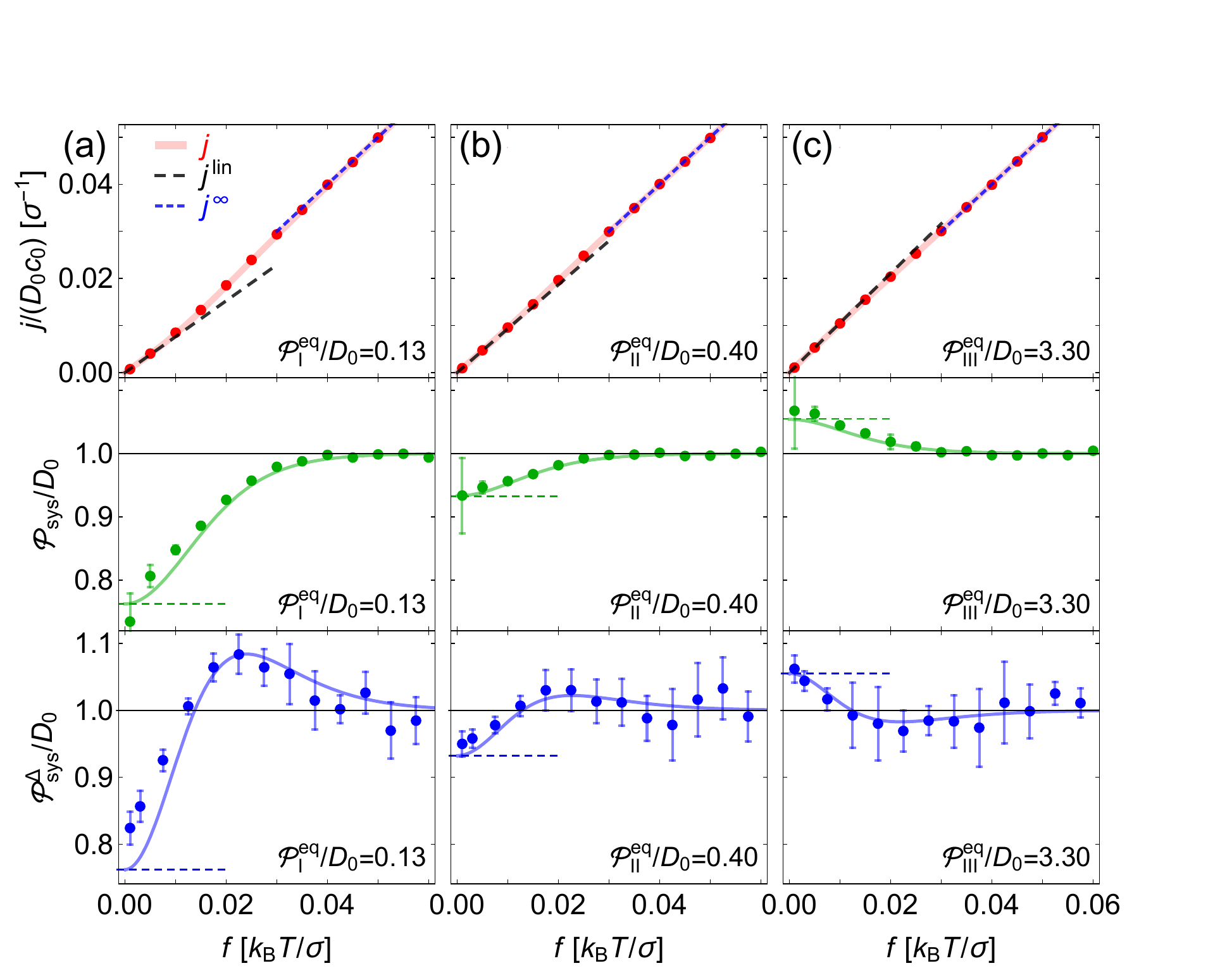}
\caption{Nonequilibrium flux and permeability results for the single-component penetrant systems, presented for the three penetrant types I-III (cf. Table.~I) of different equilibrium membrane permeabilities \figidx{a}~$\mathcal{P}^\text{eq}_\text{I}/D_0=0.13$, \figidx{b}~$\mathcal{P}^\text{eq}_\text{II}/D_0=0.4$, and \figidx{c}~$\mathcal{P}^\text{eq}_\text{III}/D_0=3.3$. Top panels: steady-state flux $j(f)/(D_0 c_0)$ from the exact solution [red solid lines, see \Eq~\eqref{eq:j2}], the leading-order expression $j^\text{lin}$ [black dashed lines, see \Eq~\eqref{eq:j3}], and $j^\infty$ [blue dashed lines, see \Eq~\eqref{eq:j_f_infty}], compared with the simulation results (symbols).
Middle panels: system permeability, $\mathcal{P}_\text{sys} = j(f)/(c_0\beta f)$, normalized by $D_0$, obtained from the theory [solid lines, see \Eq~\eqref{eq:forcedpenP}] and simulations (symbols). 
Bottom panels: differential system permeability, $\mathcal{P}^{\Delta}_\text{sys} = dj(f)/(c_0\beta df)$ normalized by $D_0$, obtained from the theory (solid lines) and simulations (symbols). The horizontal dashed lines depict the limiting values of $\mathcal{P}_\text{sys}^\text{eq}$ for $f\rightarrow 0$ [see \Eq~\eqref{eq:forcedpenP_linear}]. The horizontal solid lines depict $\mathcal{P}_\text{sys} /D_0 =\mathcal{P}^{\Delta}_\text{sys} /D_0 =1$. The used parameter values are summarized in \Tab~\ref{tab:para}.
}
\label{fig:j_1comp}
\end{figure*}

Limiting expressions for the concentration $c(z,f)$ reduce to
\begin{eqnarray}\label{eq:cf0}
c(z)=
\begin{cases}
{c_0} \mathcal{K} &z_\text{L} \leq z \leq z_\text{R}, \\
{c_0} &\text{elsewhere},
\end{cases}
\end{eqnarray}
not only for $f \rightarrow 0$ (see the gray arrows in \Fig~\ref{fig:3}) but also when the membrane is as permeable as the bulk ($\mathcal{P}^\text{eq} = D_0$).
In addition, we find a notable limiting expression for the central concentration
\begin{equation}\label{eq:cfinf}
\lim_{f \rightarrow \infty}\frac{c_\text{in}(L/2)}{c_0} = \frac{D_0}{D_\text{in}},
\end{equation}
where the penetrant's membrane diffusivity in equilibrium can be estimated by measuring the nonequilibrium concentration ratio under a high force driving (see the red arrows in \Fig~\ref{fig:3}). 

Overall, the simple continuum level Smoluchowski picture agrees very well qualitatively and even semi-quantitatively with the simulation results for $c(z,f)$.
However, the theory exhibits limitations. The quantitative deviations from the simulation results can be attributed to the complexity arising from the polymer network membrane. The membrane is not a simple homogeneous medium. In stark contrast, it exhibits force- and location-dependent (volume) responses, especially with high forces and strong attractions.  Moreover, the simulated membrane-bulk system does not provide a piecewise constant landscape [$G(z), D(z)$] as ideally imposed in the theory, leading to the largest differences between the two approaches at the interfaces. Nevertheless, our simple theory is qualitatively valid for the force range we consider in this study, which in principle enables us to quantify $\mathcal{P}^\text{eq}$ by analyzing $c(z,f)$.

\subsubsection{Force-dependent flux and system permeabilities}

Now we focus on the steady-state flux $j$, obtained in a closed-form as a nonlinear function of $f$ (see Appendix for derivation),  
\begin{equation}
\label{eq:j2}
j = {  D_0 c_\text{0}  \beta f}\left[ 1+ {\left(\frac{D_0}{\mathcal{P}^\text{eq}}-1 \right) S(f)}\right]^{-1},
\end{equation}
where $S(f) \equiv {\sinh \left({\beta  f d}/{2}\right)}/{\text{sinh}\left({\beta  f L}/{2}\right)}$.
The flux has no zeroth order term of $f$ [cf. \Eq~\eqref{eq:j_example2}], and the leading-order term for small force $f~\ll~\kT / L \ll \kT / d \ll \kT / \sigma$ yields
\begin{eqnarray}
\label{eq:j3}
j^\text{lin} &=&  D_0 c_\text{0} \beta f \left[ 1+ \left( \frac{D_0}{\mathcal{P}^\text{eq}} -1\right) \frac{d}{L} \right]^{-1} ,
\end{eqnarray}
solely described by $\mathcal{P}^\text{eq}$, $c_0$, $D_0$, and the rescaled membrane thickness $d/L$, which are all equilibrium quantities.

Another important feature of $j$ is found from the high force regime, where the membrane as an energy barrier becomes negligible. The flux simply reduces to the relation for the homogeneous bulk solution
\begin{eqnarray}
\label{eq:j_f_infty}
j^{\infty}=D_0 c_\text{0}  \beta f,
\end{eqnarray}
implying that the nonlinear nature of the flux $j$ lies on an intermediate range of the force, $ \kT/L \leq f \leq \kT/d$ ($0 \lesssim f \lesssim 0.06~\kT/\sigma$), otherwise the flux in the both infinitely small and large limits is simply the linear function of $f$, \ie, $j^\text{lin}$ or $j^{\infty}$.

In the top panels of \Fig~\ref{fig:j_1comp} the exact expression for the flux $j$ [red solid lines, \Eq~\eqref{eq:j2}], rescaled by $D_0c_0$, is compared with the leading-order expression $j^\text{lin}$ [\Eq~\eqref{eq:j3}] and $j^{\infty}$ [\Eq~\eqref{eq:j_f_infty}] as well as simulation results for $j=\langle c(z) v_z(z)\rangle$ (symbols), for different $\mathcal{P}^\text{eq}$. The standard error bars are smaller than the symbol size.
The theoretical prediction for $j$ is in excellent agreement with the simulation results.
The nonlinearity of $j$ is more significant for smaller $\mathcal{P}^\text{eq}$ [\Fig~\ref{fig:j_1comp}\figidx{a}]. 
For $\mathcal{P}^\text{eq} \gg D_0$ the flux is almost linear, as shown in the top panels of \Figs~\ref{fig:j_1comp}\figidx{b} and \figidx{c}.

\begin{figure*}[t]
\centering
\includegraphics[width = 1.02\linewidth]{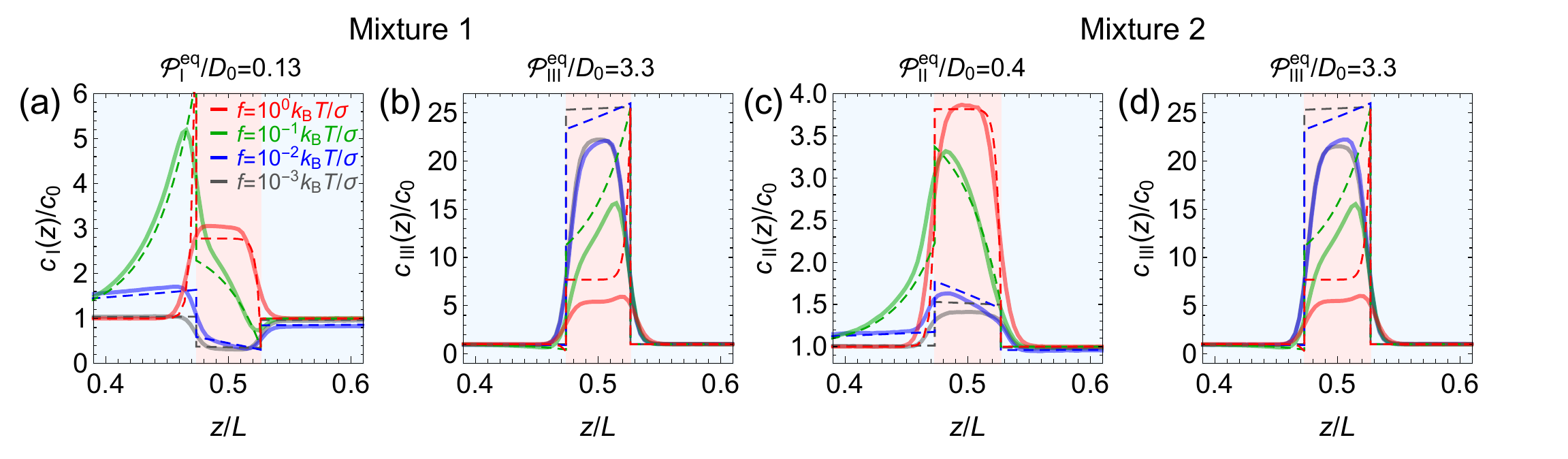}
\caption{Penetrant concentration profiles in two-component penetrant mixtures, (mixture 1 of $\{\text{p}_\text{I}, \text{p}_\text{III}\}$, and mixture 2 of $\{\text{p}_\text{II}, \text{p}_\text{III}\}$), presented for different forces. The red shaded areas depict the membrane regions with equilibrium thicknesses $d/\sigma=16$ and $d/\sigma=16.4$, for mixtures 1 and 2, respectively. The simulation results (solid lines) and the theoretical prediction (dashed lines) from the steady-state Smoluchowski solution [\Eq~\eqref{eq:sol_full}] are compared for different parameters for mixture 1 with \figidx{a}~$\{\beta\epsilon_\text{npI} = 0.1, \mathcal{P}_\text{I}^\text{eq}/D_0=0.13 \}$ and \figidx{b}~$\{\beta\epsilon_\text{npIII} = 1.2, \mathcal{P}_\text{III}^\text{eq}/D_0=3.3\}$, and for mixture 2 with \figidx{c}~$\{\beta \epsilon_\text{npII} = 0.6, \mathcal{P}_\text{II}^\text{eq}/D_0=0.4 \}$ and \figidx{d}~$\{\beta\epsilon_\text{npIII} = 1.2, \mathcal{P}_\text{III}^\text{eq}/D_0=3.3\}$. The used parameter values are summarized in \Tab~\ref{tab:para}.
}
\label{fig:cz_twocomp}
\end{figure*}

\begin{figure}[h]
\centering
\includegraphics[width = 1.1\linewidth]{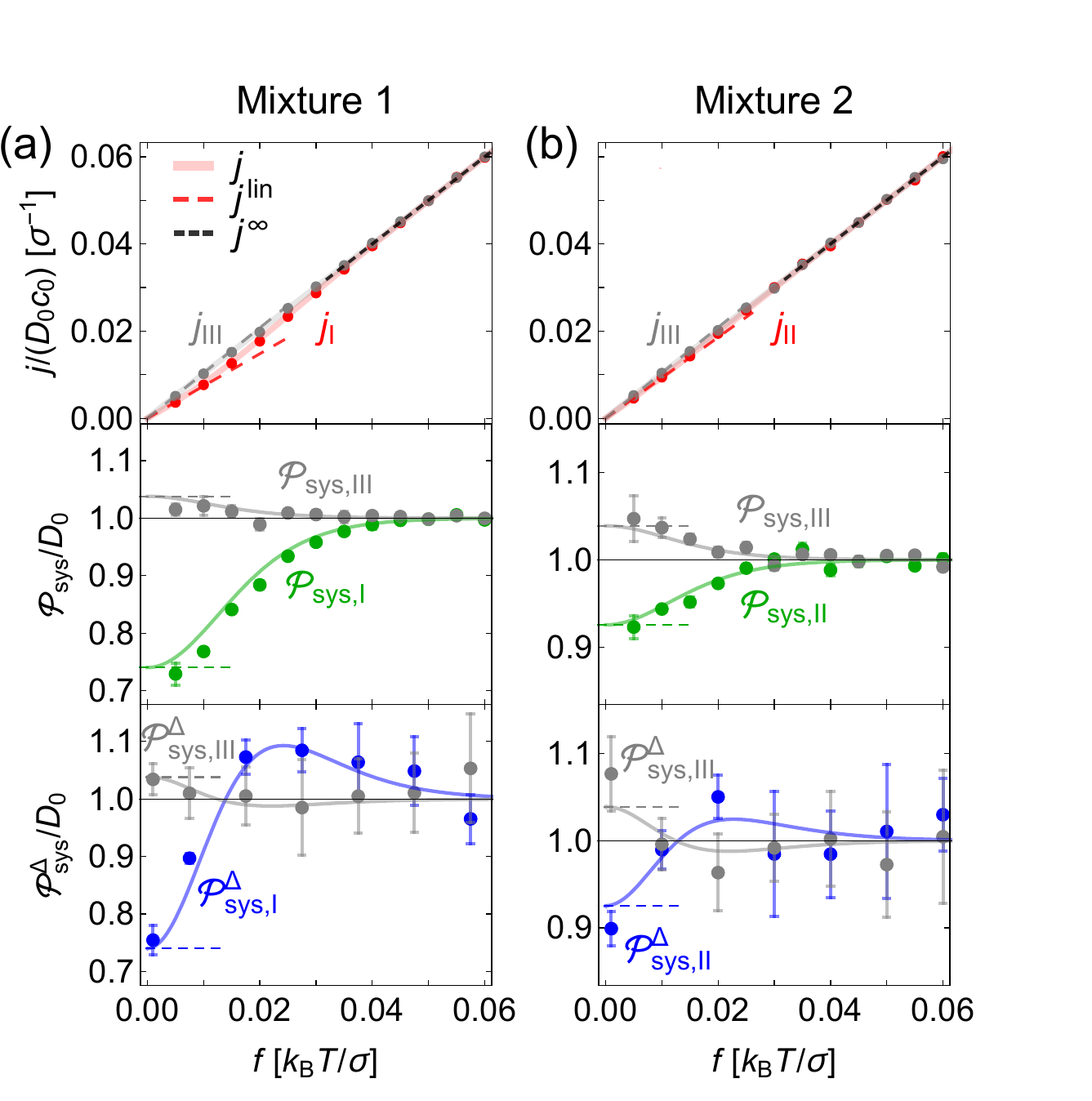}
\caption{Results from the two-component penetrant mixtures, for \figidx{a}~mixture 1: $\mathcal{P}_\text{I}^\text{eq}/D_0=0.13,~\mathcal{P}_\text{III}^\text{eq}/D_0=3.3$, and \figidx{b}~mixture 2: $\mathcal{P}_\text{II}^\text{eq}/D_0=0.4,~\mathcal{P}_\text{III}^\text{eq}/D_0=3.3$. Top panels: steady-state flux $j(f)/(D_0 c_0)$ from the exact solution [red and gray solid lines, see \Eq~\eqref{eq:j2}], $j^\text{lin}$ (red and gray dashed lines), and $j^\infty$ (black dashed lines), compared with the simulation results (symbols).
Middle panels: system permeability $\mathcal{P}_\text{sys} = j(f)/(c_0\beta f)$ normalized by $D_0$, obtained from the theory [solid lines, see \Eq~\eqref{eq:forcedpenP}] and simulations (symbols).
Bottom panels: differential system permeability, $\mathcal{P}^{\Delta}_\text{sys} = dj(f)/(c_0\beta df)$ normalized by $D_0$, obtained from the theory (solid lines) and simulations (symbols). The horizontal dashed lines depict the limiting values of $\mathcal{P}_\text{sys}^\text{eq}$ for $f\rightarrow 0$ [see \Eq~\eqref{eq:forcedpenP_linear}]. The horizontal solid lines depict $\mathcal{P}_\text{sys} /D_0 =\mathcal{P}^{\Delta}_\text{sys} /D_0 =1$. The used parameter values are summarized in \Tab~\ref{tab:para}.}
\label{fig:j_2comp}
\end{figure}

In fact, as we discussed in Sec.~B.3, the steady-state flux obtained in \Eq~\eqref{eq:j2} is the system's flux embracing the bulk reservoir and the membrane.
Using our definition, \Eq~\eqref{eq:P_j}, we obtain the nonlinear nonequilibrium system permeability, 
\begin{equation}
\label{eq:forcedpenP}
 \mathcal{P}_\text{sys}(f) = { D_0 }\left[1+{\left(\frac{D_0}{\mathcal{P}^\text{eq}}-1 \right)S(f)}\right]^{-1}.
\end{equation}
In the limit of $f \rightarrow 0$, we find that $\mathcal{P}_\text{sys}$ reduces to the proportionality constant of the foregoing linear response $j^\text{lin} = \mathcal{P}_\text{sys}^\text{eq} c_\text{0} \beta f$ as
\begin{eqnarray}
\label{eq:forcedpenP_linear}
 \mathcal{P}_\text{sys}^\text{eq} = D_0 \left[ 1+ \left( \frac{D_0}{\mathcal{P}^\text{eq}} -1\right) \frac{d}{L} \right]^{-1},
\end{eqnarray}
which provides a relation between system permeability and membrane permeability in equilibrium. Rewritten, it reveals a known reciprocal summation rule for the equilibrium system permeability~\cite{diamond1974interpretation},
\begin{eqnarray}
\label{eq:Precp}
\frac{L}{\mathcal{P}_\text{sys}^\text{eq}}&=&\frac{(L-d)/2}{\mathcal{P}_\text{0L}^\text{eq}} + \frac{d}{\mathcal{P}^\text{eq}} + \frac{(L-d)/2}{\mathcal{P}_\text{0R}^\text{eq}},
\end{eqnarray}
with the equilibrium bulk permeability denoted by $\mathcal{P}_\text{0L}^\text{eq}=\mathcal{P}_\text{0R}^\text{eq}=D_0$. 
In addition, in the limit of $f \rightarrow \infty$, we find $\mathcal{P}_\text{sys}^\infty = D_0$

In the middle horizontal panels of \Fig~\ref{fig:j_1comp}, the theoretical prediction for $\mathcal{P}_\text{sys}(f)$ obtained in \Eq~\eqref{eq:forcedpenP} is depicted by the solid line, showing a very good agreement with the simulation results analyzed from $\langle c(z,f) v_z(z,f)\rangle/(c_0 \beta f)$ (symbols). 
The system permeability, $\mathcal{P}_\text{sys}(f)= j / (c_0 \beta f)$, monotonically increases or decreases with $f$ from the equilibrium value ($\mathcal{P}_\text{sys}^\text{eq}$) to the limiting value $\mathcal{P}_\text{sys}^\infty = D_0$.
This change of $\mathcal{P}_\text{sys}(f)$, signifying the nonlinearity of $j$, indeed mostly occurs in the range of $ \kT/L \leq f \leq \kT/d$.

In addition, we find a closed-form solution for the nonequilibrium differential system permeability via our definition $\mathcal{P}^{\Delta}_\text{sys}(f)=d j / (c_0 \beta df)$ [see \Eq~\eqref{Aeq:p_inc} in Appendix for the full closed-from expression].

In the bottom panels of \Fig~\ref{fig:j_1comp}, the analytical results for $\mathcal{P}^{\Delta}_\text{sys}(f)$ are compared with the discrete derivatives of the simulation data for $j$.
The major characteristics of $\mathcal{P}^{\Delta}_\text{sys}$, \ie, a {\it differential} response function of $f$, reveals that the nonlinear response of the flux is a dramatically varying nonmonotonic function of $f$, particularly for less permeable systems. This reflects that the penetrants' response in terms of the permeability can be largely controlled by external driving forces.

We note here that our finding of the nonmonotonic nature of $\mathcal{P}^{\Delta}_\text{sys}(f)$ bears a resemblance to the effective diffusivity found in ideal 1D tilted periodic potentials~\cite{costantini1999threshold,reimann2001giant,burada2009diffusion,hanggi2009artificial}. It turned out that a small free diffusion coefficient of a particle (e.g., particles with high friction) can trigger a large enhancement of the particle's mobility in washboard-like potentials when optimally tilted. This tendency is similar to our finding that the maximum of the differential system permeability is enhanced with the lower equilibrium membrane permeability, but in our case $D_0$ is fixed.
However, contrary to the idealized washboard model, our theoretical model specifically focuses on transport in membranes in the language of permselectivity, which is compared with the polymer-resolved membrane simulations.   

More discussion on the nonmonotonicity of $\mathcal{P}^{\Delta}_\text{sys}(f)$ is worthwhile in connection with the penetrant concentration profile. 
From the results in \Figs~\ref{fig:j_1comp}\figidx{a} and \figidx{b}, it turns out that the intermediate force $\beta f=0.1/\sigma$, which induces the notable cis-side accumulation of the penetrants [see the green lines in \Figs~\ref{fig:3}\figidx{a} and \figidx{b}], is actually high enough for the sake of gaining the linear flux ($j^{\infty}$) and thus converging $\mathcal{P}^{\Delta}_\text{sys}=D_0$.
The maximal response of $\mathcal{P}^{\Delta}_\text{sys}(f)$ arises at around $\beta f=0.02/\sigma$ [see  the peaks in the bottom panels of \Figs~\ref{fig:j_1comp}\figidx{a} and \figidx{b}].
At this optimal force, the penetrants accumulate on the cis-side interface, which builds asymmetry between $c_\text{0L}(z,f)$ and $c_\text{0R}(z,f)$ across the membrane toward the trans side, resulting in the gradually decreasing  concentration gradient [see the blue lines in \Figs~\ref{fig:3}\figidx{a} and \figidx{b}].
This balance between the bulk concentration asymmetry and the moderate linear gradient of the inner-membrane concentration [close to the Fick type permeation shown in \Fig~\ref{fig:models}\figidx{c}] is sensitively modulated by changing the force by a small amount, which tunes the permeability. 
In this force regime, the differential system permeability is maximized assisted by the both driving force ($f$) and the concentration difference ($\Delta c$) in a cooperating fashion [see \Eq~\eqref{eq:j_example2} and subsequent discussions].
It is interesting to observe the opposite feature in \Fig~\ref{fig:j_1comp}\figidx{c}: $\mathcal{P}^{\Delta}_\text{sys}(f)$ is minimized at around almost the same small optimal force ($\beta f=0.02/\sigma$) at which the gradual increase of $c_\text{in}$ occurs [see the blue line in \Fig~\ref{fig:3}\figidx{c}].
Thus, with the higher $\mathcal{P}^\text{eq}$ and the intermediate force, the differential system permeability is maximally hindered by the backward flux originated from the Fick type mechanism dominated by the concentration gradient.

\subsection{Mixture of two-component penetrants}
\label{sec:twocomp}

Now we turn to two-component mixtures of penetrant types $\text{p}_\text{I}$, $\text{p}_\text{II}$, and $\text{p}_\text{III}$. 
In mixture 1 of $\{\text{p}_\text{I},\text{p}_\text{III}\}$, $\text{p}_\text{I}$ is repulsive to the membrane and $\text{p}_\text{III}$ is strongly attractive. 
In mixture 2 of $\{\text{p}_\text{II},\text{p}_\text{III}\}$, $\text{p}_\text{II}$ is moderately attractive to the membrane and $\text{p}_\text{III}$ is strongly attractive. 
For all cases, the penetrant-penetrant interaction is always repulsive (steric), as in the single-component systems. 

In general, such a binary mixture tends to exhibit a more intricate nature than the single-component system, owing to potentially inherent coupled interactions mediated by the membrane and many-body effects~\cite{nguyen2019many}, especially when some of the immanent interactions are strong. The developed theory in this work is based on single-component penetrants in the system, and thus is not strictly applicable to multi-component mixtures. However, when the above-mentioned coupled interactions and many-body effects are negligibly small, one may expect that the single-component Smoluchowski solutions are additive.

We aim to examine the superposition principle and decomposition of our theory throughout the simulated binary mixture within the given parameter sets.
Another important objective in this section, which we attempt to achieve within our theoretical framework, is to explore the feature of force-dependent selective permeation in a mixture of different penetrants.
We look for a major role of external driving forces that control the permselectivity in the mixture, in which we simulate penetrants of different types in the presence of a single membrane type.

\subsubsection{Penetrant concentration profiles}

In \Fig~\ref{fig:cz_twocomp}, the simulation results for the penetrant concentration profiles $c_\text{I}(z)$ and $c_\text{III}(z)$ in each mixture are compared with the theoretical prediction [\Eq~\eqref{eq:sol_full}]. The results exemplify that even for the mixtures the decomposed Smoluchowski picture applies, in which we consider only mutually repulsive penetrants and the equilibrium membrane permeability up to $\mathcal{P}^\text{eq}_\text{III}\approx 3~D_0$. But we also observe a limitation of the theory, especially for the highly attractive penetrant ($p_\text{III}$), which deviates more from the simulation results in comparison with the single-component cases.
However, the main characteristics of the gradient change of $c_\text{in}(z,f)$ depending on $\mathcal{P}^\text{eq}$ with the corresponding threshold argument, $\mathcal{P}^\text{eq}=D_0$, are still validly captured from the theoretical prediction.
The discrepancy particularly found in \Figs~\ref{fig:cz_twocomp}\figidx{b} and \figidx{d} is expected because the polymer membrane system involves additional complex features in membrane responses and many-body correlations, which are not accounted for in the theory. 
Nevertheless, the result shows that our simple theory can also be utilized to characterize such cosolute mixtures, especially with low intrinsic (equilibrium) membrane permeabilities.

\subsubsection{Force-dependent flux and system permeabilities}

We show the analytical results for the flux, the system permeability, and the differential system permeability in \Fig~\ref{fig:j_2comp}, compared with the simulation results.
The fluxes in the mixtures exhibit overall a similar tendency than observed in the single-component system, in which the nonlinearity with respect to $f$ is more significant for intrinsically less permeable penetrants, $\text{p}_\text{I}$.
The comparison of $j$ demonstrates the validity of the prediction, \Eq~\eqref{eq:j2}, for the two-component mixture, which even quantitatively matches with the simulation results, demonstrating the additivity within the chosen parameter sets.
The system permeabilities (middle horizontal panels in \Fig~\ref{fig:j_2comp}) first start at the equilibrium values ($\mathcal{P}_\text{sys,I}^\text{eq}$ and $\mathcal{P}_\text{sys,III}^\text{eq}$) for very small forces and gradually and monotonically converge to the bulk permeability $D_0$ for very large forces, which is similar as separately found for the single-component system in \Fig~\ref{fig:j_1comp}. Therefore, we confirm that this tendency for each permeability in the same mixture is effectively captured by our theory in a decomposed fashion.
The predicted differential system permeabilities $\mathcal{P}_\text{sys}^\Delta(f)$ (solid lines in bottom panels) also match well with the simulation results.
The observed simultaneous maximization and minimization of $\mathcal{P}_\text{sys}^\Delta(f)$ for the different penetrant types in the mixture reflects the immanence of selectivity that can be largely controlled by a small changes of the driving force, particularly in the mixture with low permeability.

In the shown range of the force, $0<f \lesssim 0.06~\kT/\sigma$, $\mathcal{P}_\text{sys}$ and $\mathcal{P}_\text{sys}^\Delta$ converge to $D_0$, which is associated with the membrane width via $\kT/L \leq f \leq \kT/d$: The used parameter value $d \approx 16\sigma$ corresponds to $f\approx 0.06~\kT/\sigma$.

\subsubsection{System selectivity}

The system selectivity, $\alpha_\text{sys} \equiv \mathcal{P}_\text{sys,III}(f)/\mathcal{P}_\text{sys,I}(f)$, which extends the original definition for equilibrium membrane selectivity~\cite{park2017} to the system's nonequilibrium selectivity, is thus the ratio between fluxes and bulk concentrations,
\begin{eqnarray}\label{eq:a}
\label{eq:alpha_tot}
\alpha_\text{sys}=\frac{j_\text{III}/c_{0,\text{III}} }{j_\text{I} /c_{0,\text{I}}},
\end{eqnarray}
whereas the nonequilibrium differential system selectivity $\alpha^{\Delta}_\text{sys}(f) \equiv \mathcal{P}^{\Delta}_\text{sys,III}(f) / \mathcal{P}^{\Delta}_\text{sys,I}(f)$ is
\begin{eqnarray}\label{eq:Da}
\label{eq:alpha_tot}
\alpha^{\Delta}_\text{sys}=\frac{j_\text{III}'/c_{0,\text{III}} }{j_\text{I}' /c_{0,\text{I}}},
\end{eqnarray}
with $j' \equiv dj/df$. The same definition applies to mixture 2 by replacing I with II.

\begin{figure}[t]
\centering
\includegraphics[width = 1\linewidth]{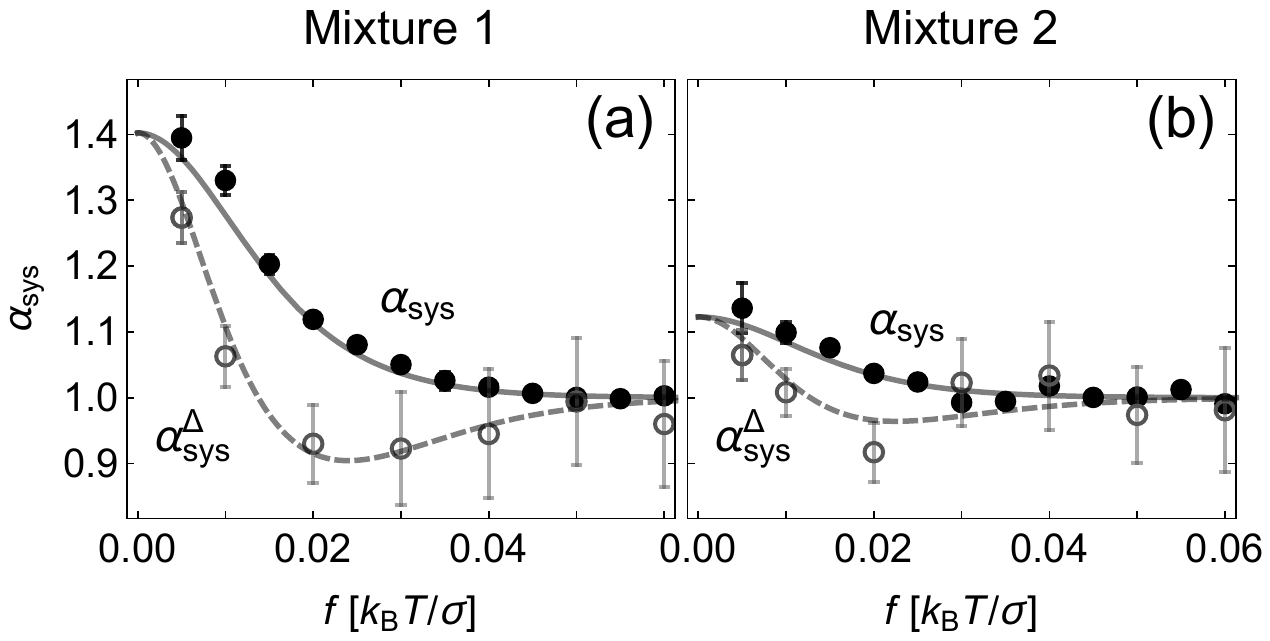}
\caption{System selectivity, $\alpha_\text{sys}(f)$, in the two-component penetrant mixtures, for \figidx{a}~mixture 1: $\mathcal{P}_\text{I}^\text{eq}/D_0=0.13,~\mathcal{P}_\text{III}^\text{eq}/D_0=3.3$, and \figidx{b}~mixture 2: $\mathcal{P}_\text{II}^\text{eq}/D_0=0.4,~\mathcal{P}_\text{III}^\text{eq}/D_0=3.3$. The system selectivities $\alpha_\text{sys} = \mathcal{P}_\text{sys,III}(f)/\mathcal{P}_\text{sys,I}(f)$ and $\alpha_\text{sys} = \mathcal{P}_\text{sys,III}(f)/\mathcal{P}_\text{sys,II}(f)$ [solid lines, see \Eq~\eqref{eq:a}] are compared with the simulation results (filled circles). The differential system selectivities $\alpha^{\Delta}_\text{sys}(f) = \mathcal{P}^{\Delta}_\text{sys,III}(f)/\mathcal{P}^{\Delta}_\text{sys,I}(f)$ and $\alpha^{\Delta}_\text{sys}(f) = \mathcal{P}^{\Delta}_\text{sys,III}(f)/\mathcal{P}^{\Delta}_\text{sys,II}(f)$ are depicted by the dashed line [see \Eq~\eqref{eq:Da}], compared with the simulation results (empty circles).}
\label{fig:a_21}
\end{figure}

In \Figs~\ref{fig:a_21}\figidx{a} and \figidx{b}, we show the system selectivity $\alpha_\text{sys}(f)$ (solid lines) and the differential system selectivity $\alpha^{\Delta}_\text{sys}(f)$ (dashed lines) for mixtures 1 and 2, respectively. The simulation results for $\alpha_\text{sys}(f)$, depicted by symbols, are compared with the analytical results based on \Eq~\eqref{eq:j2}.
For high forces, both system and differential system selectivities converge to unity, meaning that there is no selectivity and that all penetrants flow with the flux $j^{\infty}$ derived in \Eq~\eqref{eq:j_f_infty}, dictated by the exceedingly high force.
However, in a small force range in which $j$ is predominantly nonlinear, the selectivity is most sensitive to the force, which bears a potential fine-tuning of the permselectivity.
Note that we observe an undulation in $\alpha^{\Delta}_\text{sys}(f)$ that crosses over the unity at around $\beta f=0.01/\sigma$.
Thus, the differential selectivity can be well controlled by changing the force by a small amount.

\subsection{Intrinsic membrane permeability and selectivity}

So far, we focussed on the permeability and the differential permeability of the whole membrane-reservoir system. However, how to quantify the intrinsic membrane permeability therein (i.e. independent from system size and boundary conditions, as in equilibrium) and its selectivity is another important quest, if possible at all. 

It is not so straightforward to obtain the force-dependent membrane permeability, $\mathcal{P}(f)$,  because there is no unequivocal definition.
Nevertheless, one may employ the reciprocal summation rule [\Eq~\eqref{eq:Precp}] in order to obtain the membrane permeability, at least for systems close to equilibrium ($f \ll \kT/L$).
To this end, we rewrite $L/\mathcal{P}_\text{sys}(f)$ using \Eq~\eqref{eq:forcedpenP} as the partial fraction decomposition 
\begin{equation}
\label{eq:L/Ptot}
\frac{L}{\mathcal{P}_\text{sys}}=\frac{L}{D_0}\left[1- S(f)\right]+ \frac{L}{\mathcal{P^\text{eq}}} S(f),
\end{equation}
with $\mathcal{P}_\text{sys}=j/(c_0\beta f)$ [see \Eq~\eqref{eq:P_j}].

In the limit $f \rightarrow 0$, the first term in \Eq~\eqref{eq:L/Ptot} reduces to $(L-d)/D_0$ and the second term reduces to $d/\mathcal{P}^\text{eq}$, from which we conclude that the force-dependent (chordal) membrane permeability is
\begin{eqnarray}\label{eq:Pinf}
\mathcal{P}(f) 
                                     &=& \mathcal{P}^\text{eq}\frac{d}{L} \frac{\sinh \left({\beta  f L}/{2}\right)}{ \sinh\left({\beta  f d}/{2}\right)},
\end{eqnarray}
and we find the $\mathcal{P}$-$j$ relation using \Eqs~\eqref{eq:Pinf} and \eqref{eq:j2},
\begin{eqnarray}\label{Pinf_j}
\mathcal{P}(j) = \frac{j\left(\mathcal{P}^\text{eq}-D_0\right)}{j-j^\infty}\frac{d}{L}. \nonumber
\end{eqnarray}
The leading-order expression of $\mathcal{P}(f)$ for small $f$ reads
 \begin{eqnarray}
\label{eq:forcedpenPin_low}
 \mathcal{P} \approx  \mathcal{P}^\text{eq} \left[ 1 +  \frac{L^2-d^2}{24} \left(\beta f\right)^2\right].  
\end{eqnarray}
As shown in \Fig~\ref{fig:P_alpha}, the membrane permeabilities $\mathcal{P}(f)$
first start at equilibrium values ($\mathcal{P}_\text{I}^\text{eq}$ and $\mathcal{P}_\text{III}^\text{eq}$) and then monotonically and rapidly increase with $f$, as is the feature of \Eq~\eqref{eq:Pinf}.
This diverging nature of $\mathcal{P}(f)$ in the range of $0< f \lesssim 0.05~\kT/\sigma$ is indeed responsible for the convergence of the system permeability $\mathcal{P}_\text{sys}(f)$ shown in \Fig~\ref{fig:j_2comp}, as a consequence of the reciprocal summation rule $L/\mathcal{P}_\text{sys} = d/\mathcal{P} + C$, where $C=(L-d)/D_0$ becomes a constant for large $f$. 
The dashed lines depict the leading-order expressions obtained in \Eq~\eqref{eq:forcedpenPin_low}.
However, as it turns out by the comparison with simulation results (symbols in insets), this approach using the reciprocal summation rule marginally agrees only in a very small force regime. 

\begin{figure}[t]
\centering
\includegraphics[width = 1\linewidth]{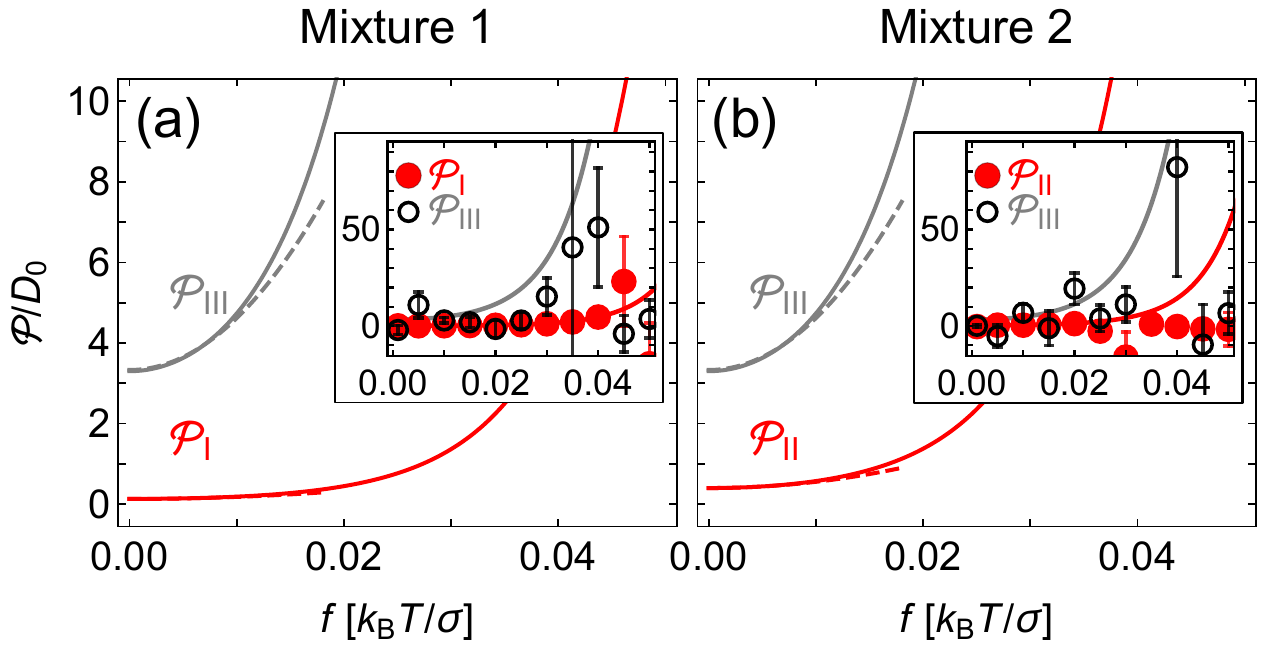}
\caption{Membrane permeabilities, $\mathcal{P}(f)/D_0$ \Eq~\eqref{eq:Pinf}, from the two-component mixtures, for \figidx{a}~mixture 1: $\mathcal{P}_\text{I}^\text{eq}/D_0=0.13,~\mathcal{P}_\text{III}^\text{eq}/D_0=3.3$, and \figidx{b}~mixture 2: $\mathcal{P}_\text{II}^\text{eq}/D_0=0.4,~\mathcal{P}_\text{III}^\text{eq}/D_0=3.3$. The solid lines depict the prediction presented in \Eq~\eqref{eq:Pinf}.
 The dashed lines depict the leading-order expressions obtained in \Eq~\eqref{eq:forcedpenPin_low}. Insets show $\mathcal{P}/D_0$ with wider $y$-axis range, in comparison with simulation results depicted by symbols.}
\label{fig:P_alpha}
\end{figure}

Finally, the membrane selectivity $\alpha \equiv \mathcal{P}_\text{III}/\mathcal{P}_\text{I}$ is then found according to \Eq~\eqref{eq:Pinf} as $\alpha = \mathcal{P}_\text{III}^\text{eq}/\mathcal{P}_\text{I}^\text{eq}$,
\ie, the ratio between the equilibrium membrane permeabilities, which turns out to be force-independent in this particular definition of force-dependent membrane permeability. This is in stark contrast to the system selectivities shown in \Fig~\ref{fig:a_21}. 
Note that the results obtained here are derived based on the reciprocal rule in equilibrium. The results might reflect that an intrinsic membrane permeability can not be defined in a universal fashion and might be context-independent, i.e., not simply additive superpositions. This may not come as a surprise because this is a general consequence of nonlinear systems.

\section{Conclusion}

We have investigated nonequilibrium permeability and selectivity for solute transport across a polymer membrane under the action of a uniform driving force, using penetrant- and polymer-resolved simulation and Smoluchowski theory. Thereby, we have gone beyond the `classical' linear response solution-diffusion model and extended concepts and definitions of permeability towards nonequilibrium situations. In particular, we have presented and discussed possible definitions of a force-dependent system permeability and permselectivity by analytical solutions of the Smoluchowski equation in a steady state, and have verified them by simulations.  As an important consequence, we have demonstrated that the solute selectivity of the system can be tuned and controlled by a small change of the external force. This force-control becomes more powerful in low-permeable systems with salient nonlinearity.

Our study broadens the fundamental understanding of permeability in nonequilibrium transport and provide a foundation developing more theoretical tools to measure and interpret the permeability in driven transport in applications, in particular in strongly driven situations beyond linear response.  This may include molecular sieving, translocation, and separation in electrophoresis~\cite{hjerten1963molecular,ajdari1991free,nkodo2001diffusion}, transport across a membrane controlled by mechanically driven forces~\cite{alberts2002carrier}, and reverse osmosis or (electro)dialysis~\cite{soltanieh1981review,mason1990statistical,reverse}, just to name a few. 

Practical situations may involve more responsive membranes, more strongly interacting solutes, and more strongly coupled hydrodynamics, and thus may require a more sophisticated model based on the presented ideas hereto.
Furthermore, the adequate definition for the nonequilibrium membrane differential permeability $\mathcal{P}^\Delta(f)$ is left as an open quest. Noteworthy is here that in nonlinear transport, the distinction between `membrane permeability' and `system permeability' might not unequivocally possible anymore. Hence, the role of system boundary conditions needs to be elucidated in more detail for nonequilibrium situations to serve for better interpretation.   Moreover, we believe that our approach will be useful in the study of transport in responsive polymer membranes~\cite{stuart2010emerging}, i.e., membranes which may deform (``open/close", even switch) locally and return feedback to the transport in the inhomogeneous, nonequilibrium flow conditions. 

\section{Methods}\label{sec:methods}

\subsection{Simulation setup and protocols}

We performed Langevin dynamics computer simulations for penetrant-membrane systems made of a polydisperse, tetra-functional polymer network including diffusive penetrants, as similarly carried out in the previous work~\cite{kim2020tuning}: We used the iteration time step $\delta \tau~=~0.005 \tau$ with the time units $\tau =~\sqrt{ {m \sigma^2} / {\kT} }$, where $m$ is the unit mass, and $\kT=1/\beta$ is the thermal energy. We used the friction coefficient $\gamma$ such that $\tau_{\gamma}=~m/\gamma = \tau$, \ie, the free penetrants' motion becomes diffusive after $200$ time steps. 
The initial configuration of the network consists of $4 \times 4 \times 4$ unit cells of a diamond cubic lattice on which cross-linkers amounting to $N_\text{xlink}=$ $64\times8=512$  are located. 
The number of monomers per chain (between two nearest neighboring cross-linkers) was randomly chosen from a uniform distribution between 2 and 18, thereby yielding on average 10 monomers per chain and mean cross-linker fraction $\approx 5\%$.
The membrane was placed at the center of the simulation box of lateral lengths $L_x = L_y = 100 \sigma$ and the longitudinal length $L \equiv L_z = 305 \sigma$, with periodic boundary conditions in all three Cartesian directions. We added $N_\text{p}=1000$ penetrant particles ($N_{\text{p}_\text{I}}=N_{\text{p}_\text{II}}=500$ for the two-component penetrants) into the bulk reservoir region, and equilibrated the whole system in the $NpT$ ensemble, such that the longitudinal box length $L$ was kept fixed, while $L_x $ and $L_y$ could adjust equivalently to fulfill the zero target pressure $p_x = p_y =0$~\cite{Erbas2015}, and the target temperature $T$. 
After equilibration for $4\times10^7$ time steps, we performed next pre-production simulations with external forces $f$ that act on the penetrants in simulation boxes with the fixed volume $V$ for $10^8$ time steps.
Finally, we performed the production run simulations for $2\times10^8$ time steps. In each time step the center of mass of the membrane (${\bf r}_\text{com}$) was calculated, by which all the membrane particles were constrained by an auxiliary tethering in terms of a harmonic force ${\bf f}_\text{com} = k_\text{com}({\bf r}_\text{center}-{\bf r}_\text{com})$.
Thus the membrane center of mass was tightly bound to the center of the simulation box (${\bf r}_\text{center}$), with the spring constant $k_\text{com} = 100 ~\kT/\sigma$ per membrane particles.
The membrane thickness $d(f)$ is computed based on the full width at half maximum~\cite{milster2021tuning}, which varies slightly with the force only for large $\mathcal{P}^\text{eq}$ [see \Fig~\ref{fig:3}\figidx{c}], otherwise it is nearly constant (see \Tab~\ref{tab:para} for the equilibrium $d$).

\subsection{Chemical force field for polymer bonds}

We employed harmonic stretching (bond) and bending (angle) potentials for the bonded interactions of our semi-flexible polymers~\cite{kim2017cosolute,kim2020tuning}. As considered in the previous work~\cite{kim2020tuning}, the bonded parameters were determined by coarse-graining from all-atom simulations of cross-linked PNIPAM-BIS (poly($N$-isopropylacrylamide)-$N,N'$-methylenebisacrylamide) chains~\cite{milster2019cross}. A cross-linker connects monomers of four polymer chains, therefore the network is tetra-functional. There are nine different (7 bending (angles), 2 stretching (bonds)) potentials in total. We determined eighteen bonded parameters $K_r^{ij}$, $r_0^{ij}$, $K_\theta^{ijk}$, and $\theta_0^{ijk}$ by fitting harmonic potentials to the free energies obtained from the all-atom simulations. The details of all bonded interactions can be found in the previous work~\cite{kim2020tuning}.

\begin{acknowledgments}
The authors thank Matthias Ballauff, Benjamin Rotenberg, Arturo Moncho-Jord{\'a} and Changbong Hyeon for fruitful discussions. 
This project has received funding from the European Research Council (ERC) under the European Union's Horizon 2020 research and innovation programme (grant agreement No.\ 646659). 
W.K.K. acknowledges the support by a KIAS Individual Grant (CG076001) at Korea Institute for Advanced Study. 
M.K. acknowledges the financial support from the Slovenian Research Agency (research core funding No.\ P1-0055). 
This work was supported by the Deutsche Forschungsgemeinschaft via the Research Unit FOR 5099 ``Reducing complexity of nonequilibrium systems.''.
The simulations were performed with resources provided by the North-German Supercomputing Alliance (HLRN). 
We thank Center for Advanced Computation at the Korea Institute for Advanced Study for providing computing resources for this work.
\end{acknowledgments}

\appendix

\begin{widetext}
\section{Flux}

For the piecewise potential and diffusivity [\Eqs~\eqref{eq:Gz} and \eqref{eq:Dz}], we obtained $I(0,L,f)$ using \Eq~\ref{eq:Iz} as,
\begin{eqnarray}\label{Aeq:I2}
I(0,L,f) &=& \left[ \int_{0}^{\frac{L}{2}-\frac{d}{2}}~dy \frac{1}{D_0} + \int_{\frac{L}{2}-\frac{d}{2}}^{\frac{L}{2}+\frac{d}{2}}~dy \frac{\enat^{\beta \Delta G}}{D_\text{in}} + \int_{\frac{L}{2}+\frac{d}{2}}^{L}~dy \frac{1}{D_0}\right] \enat^{-\beta f y} \nonumber\\
&=& -\frac{1}{\beta f}\left[ \frac{\enat^{-\beta f ({\frac{L}{2}-\frac{d}{2}})} - 1}{D_0} +  \frac{ \enat^{-\beta f ({\frac{L}{2}+\frac{d}{2}})} - \enat^{-\beta f ({\frac{L}{2}-\frac{d}{2}})} }{\mathcal{P}^\text{eq}} + \frac{\enat^{-\beta f L} - \enat^{-\beta f ({\frac{L}{2}+\frac{d}{2}})}}{D_0} \right] \nonumber \\
&=& 2 \frac{ \left(\frac{D_0 }{ \mathcal{P}^\text{eq}} -1 \right) \sinh \left(\frac{\beta  f d}{2}\right) +  \sinh \left(\frac{\beta  f L}{2}\right)}{  D_0 \beta f} \enat^{-\frac{\beta  f L}{2}},
\end{eqnarray}
thereby obtaining the flux according to \Eq~\eqref{eq:flux}
\begin{eqnarray}
\label{Aeq:j2}{
j=  \frac{  c_\text{0} D_0 \beta f}{\left(\frac{D_0}{\mathcal{P}^\text{eq}}-1 \right) \frac {\sinh \left( \beta  f d / 2\right) } { \sinh \left( \beta  f L / 2\right)}+1},
 }
\end{eqnarray}
where we eliminated $\Delta G$ and $D_\text{in}$ by using the solution-diffusion expression in equilibrium, $
\mathcal{P}^\text{eq}=D_\text{in} \enat^{-\beta \Delta G}$.

\section{Penetrant concentration profiles}

By plugging \Eq~\eqref{Aeq:I2} in \Eq~\eqref{eq:sol2}, we obtained
\begin{eqnarray}
\label{Aeq:sol_full}
\frac{c(z)}{c_0} =  \left[1 - \frac{  D_0 \beta f I(z)}{\left(\frac{D_0}{\mathcal{P}^\text{eq}}-1 \right) \frac {\sinh \left( \beta  f d / 2\right) } { \sinh \left( \beta  f L / 2\right)}+1} \right] \enat^{-\beta U(z,f)}.
\end{eqnarray}

The penetrant concentration $c(z)/c_0$ is further decomposed into three parts as,
\begin{eqnarray}
\frac{c_\text{0L}(z)}{c_0} &=& \frac{\enat^{{\beta f} z}({D_0}-{\mathcal{P}^\text{eq}}) \sinh \left(\frac{{\beta f} d}{2}\right)+{\mathcal{P}^\text{eq}} \sinh \left(\frac{{\beta f} L}{2}\right)}{({D_0}-{\mathcal{P}^\text{eq}}) \sinh \left(\frac{{\beta f} d}{2}\right)+{\mathcal{P}^\text{eq}} \sinh \left(\frac{{\beta f} L}{2}\right)},\\
\frac{c_\text{in}(z)}{c_0} &=& \enat^{-\beta \Delta G} \frac{2 (D_0-\mathcal{P}^\text{eq}) \enat^{\beta f z} \sinh \left(\frac{\beta f (d-L)}{2} \right)+D_0 \left(\enat^{\beta f L}-1\right)}{2 (D_0-\mathcal{P}^\text{eq}) \enat^{\frac{\beta f L}{2}} \sinh \left(\frac{\beta f d}{2}\right)+\mathcal{P}^\text{eq} \left(\enat^{\beta f L}-1\right)},\\
\frac{c_\text{0R}(z)}{c_0} &=& \frac{c_\text{0L}(z-L)}{c_0},
\end{eqnarray}
where we denoted $c_\text{in}(z)$ by the penetrant concentration inside the membrane $(\frac{L-d}{2} \leq z\leq \frac{L+d}{2})$, $c_\text{0L}(z)$ by the left side bulk concentration $(0 \leq z<\frac{L-d}{2})$, and $c_\text{0R}(z)$ by the right side bulk concentration $(\frac{L+d}{2}<z \leq L)$.
In the derivation, we used the equilibrium solution-diffusion membrane permeability $\mathcal{P}^\text{eq} = D_\text{in}\enat^{-\beta \Delta G}=D_\text{in} \mathcal{K}$~\cite{Rafa2017,kim2017cosolute,kim2019prl,kim2020tuning}.
The leading-order expansion leads to
\begin{eqnarray}
\frac{c_\text{0L}(z)}{c_0} =& 1+ \left[ 1+ \frac{D_\text{in}}{(D_0-D_\text{in})} \frac{L}{d} \right]^{-1}{ \beta f z } + {O}(f^2) + \cdots ,&0 \leq z < \frac{L}{2}-\frac{d}{2}, \\
\frac{c_\text{in}(z)}{c_0}  =& \frac{\mathcal{P}^\text{eq}}{D_\text{in}} + \frac{1}{2}\frac{\mathcal{P}^\text{eq}}{D_\text{in}}\frac{ (L-d) (D_0-\mathcal{P}^\text{eq})}{ d (D_0-\mathcal{P}^\text{eq})+L \mathcal{P}^\text{eq}} \beta f (L - 2 z)+ {O}(f^2) + \cdots ,&\frac{L}{2}-\frac{d}{2} \leq z \leq \frac{L}{2}+\frac{d}{2}, \\
\frac{c_\text{0R}(z)}{c_0} =& 1+ \left[ 1+ \frac{D_\text{in}}{(D_0-D_\text{in})} \frac{L}{d} \right]^{-1}{ \beta f (z-L) } + {O}(f^2) + \cdots ,&\frac{L}{2}+\frac{d}{2} < z \leq L,
\end{eqnarray}
for small $f$.

The full expression of $c_\text{in}(z)$ is
\begin{eqnarray}\label{Aeq:cin_full}
\frac{c_\text{in}(z)}{c_0}  = -\frac{{\mathcal{P}^\text{eq}} \left(D_0 \enat^{\frac{1}{2} \beta  f (d-L)}-D_0 \enat^{\frac{1}{2} \beta  f (d+L)}+({\mathcal{P}^\text{eq}}-D_0) \enat^{\beta  f (d-L+z)}+D_0 \enat^{\beta  f z}-{\mathcal{P}^\text{eq}} \enat^{\beta  f z}\right)}{D_\text{in} (D_0-{\mathcal{P}^\text{eq}}) \left(\enat^{\beta f d}-1\right)+2 D_\text{in} {\mathcal{P}^\text{eq}} \enat^{\frac{\beta f d}{2}} \sinh \left(\frac{\beta  f L}{2}\right)}. 
\end{eqnarray}

\section{Differential system permeability}

The differential system permeability, defined as $\mathcal{P}^\Delta_\text{sys}(f) \equiv \frac{1}{\beta c_0} \frac{d j(f)}{d f}$, was obtained using \Eq~\eqref{eq:j2} as
\begin{equation}\label{Aeq:p_inc}
\mathcal{P}^\Delta_\text{sys}(f) ={D_0} \mathcal{P}^\text{eq} {\frac{\left({D_0}-\mathcal{P}^\text{eq}\right) \mathcal{A}(f)+\mathcal{P}^\text{eq} \left[ \cosh (\beta f L)-1\right]}
{2 \left[ ({D_0}-\mathcal{P}^\text{eq}) \sinh \left(\frac{\beta f d}{2}\right)+\mathcal{P}^\text{eq} \sinh \left(\frac{\beta f L}{2}\right)\right]^2},
}
\end{equation}
where we introduced 
\begin{equation}\label{Aeq:p_inc2}
\mathcal{A}(f) = \beta f L \sinh \left(\frac{\beta f d}{2}\right) \cosh \left(\frac{\beta f L}{2}\right)+\sinh \left(\frac{\beta f L}{2}\right) \left\{2 \sinh \left(\frac{\beta f d}{2}\right)- \beta f d \cosh \left(\frac{\beta f d}{2}\right)\right\}.
\end{equation}
\end{widetext}


%

\end{document}